\documentclass[%
 aip,
 amsmath,amssymb,
 reprint,%
]{revtex4-1}

\usepackage{graphicx}
\usepackage{dcolumn}
\usepackage{bm}

\usepackage[utf8]{inputenc}
\usepackage[T1]{fontenc}
\usepackage{mathptmx}
\usepackage{etoolbox}

\usepackage[dvipsnames]{xcolor} 

\makeatletter
\def\@email#1#2{%
 \endgroup
 \patchcmd{\titleblock@produce}
  {\frontmatter@RRAPformat}
  {\frontmatter@RRAPformat{\produce@RRAP{*#1\href{mailto:#2}{#2}}}\frontmatter@RRAPformat}
  {}{}
}%
\makeatother
\begin{document}

\preprint{AIP/123-QED}

\title[Phase transitions in chromatin: mesoscopic and mean-field approaches
]{Phase transitions in chromatin: mesoscopic and mean-field approaches}
\author{R. Tiani}
\author{M. Jardat}%
\author{V. Dahirel}%
\affiliation{ 
Sorbonne Université, CNRS, Laboratoire PHENIX (Physicochimie des Electrolytes et Nanosystèmes Interfaciaux), 4 place Jussieu, 75005 Paris, France 
}%

\date{\today}

\begin{abstract}

By means of a minimal physical model, we investigate the interplay of two phase transitions at play in chromatin organization: (1) liquid-liquid phase separation (LLPS) within the fluid solvating chromatin, resulting in the formation of biocondensates, and (2) the coil-globule crossover of the chromatin fiber, which drives the condensation or extension of the chain. In our model, a species representing a domain of chromatin is embedded in a binary fluid. This fluid phase separates to form a droplet rich in a macromolecule (B). 
Chromatin particles are trapped in a harmonic potential to reproduce the coil and globular phases of an isolated polymer chain. We investigate the role of the droplet material B on the radius of gyration of this polymer and find that this radius varies nonmonotonically with respect to the volume fraction of B. This behavior is reminiscent of a phenomenon known as \emph{co-non-solvency}: a polymer chain in good solvent (S) may collapse when a second good solvent (here B) is added in low quantity, and expand at higher B concentration.
Additionally, the presence of finite-size effects on the coil-globule transition results in a qualitatively different impact of the droplet material on polymers of various sizes. In the context of genetic regulation, our results suggest that the size of chromatin domains and the quantity of condensate proteins are key parameters to control whether chromatin may respond to an increase of the quantity of chromatin-binding proteins by condensing or expanding. 
\end{abstract}

\maketitle

\section{\label{sec:level1} Introduction\protect\\}

Biomolecular condensates or membraneless organelles are mesoscale structures of various sizes (their radius ranges from about 10 nm to 4 $\mu$m) observed in biological cells~\cite{Brangwynne_2009, Shin_2017, Shin_2018, Gibson_2019, Weber_2019, Sabari_2020, Gouveia_2022, Rouches_2024}. These structures are usually similar to liquid or solid droplets. They appear not only in the cytosol (e.g., in the form of granules)~\cite{Brangwynne_2009, Shin_2017}, but also in the fluid contained in the nucleus of eukaryotic cells, called nucleoplasm (e.g. nucleoli or transcription factories~\cite{hnisz2017phase,Sabari_2020,Rippe_2022,wu2022modulating,mann2023transcription}). The main physical mechanism behind their formation is thought to be liquid-liquid phase separation (LLPS), a phase transition where a homogeneous liquid mixture spontaneously demixes to form distinct liquid phases~\cite{Weber_2019}. In contrast with equilibrium systems, the growth of separated liquid domains by Oswald ripening is arrested in biological cells. Some authors have suggested that chemical reactions may regulate droplet size~\cite{Weber_2019,Zwicker2022}. The formation of such condensates by LLPS arises from weak, multivalent and attractive interactions between biopolymers, in particular Intrinsically Disordered Proteins (IDPs)~\cite{Shin_2018}. Their dysregulation or their related liquid-to-solid phase transitions might lead to pathological protein aggregation related to diseases~\cite{wang2021liquid}.

In the cell nucleus, condensates are formed by LLPS of nucleoplasm RNA and protein factors~\cite{hnisz2017phase,Sabari_2020}. Many nuclear regulatory processes involve the formation of condensates : DNA replication and repair~\cite{Fernandez_2023}, transcription~\cite{mann2023transcription}, and RNA processing~\cite{Shin_2018, Sabari_2020}. 
These functions are also known to be related to the state of the chromatin fiber, and in particular its level of local compaction. 
Chromatin is a long heteropolymer made of DNA associated with histone proteins.
Chromatin is organized in \textit{epigenetic} (genomic) domains or blocks~\cite{Filion2010,Boettiger2016, Cattoni2017,Szabo2018}, with privileged intra-domain contacts quantified through Chromosome Capture techniques~\cite{Sexton2012}.
The interplay between the folding dynamics of such domains and the presence of nuclear condensates may be of crucial importance to understand the physical and chemical regulation of genetic metabolism. 

Despite many experimental and modeling studies of chromatin organization and nuclear condensates, the coupling mechanisms of epigenetic domains and LLPS-driven droplets are yet unclear. From a physicochemical point of view, an epigenetic domain can be viewed as a polymer chain. In particular, block copolymer models have been proposed for \emph{Drosophila} chromosomes: each block corresponds to an epigenetic domain and each monomer interacts preferentially with other monomers of the same epigenetic type \cite{Jost2014}. Some authors have suggested that chromatin domains might experience a classical phase transition of polymer chains, the coil-globule transition between two configurational states, a swollen coiled state and a collapsed globular state~\cite{grosberggiant,Nishio79,Grassberger1995}. This transition is governed by the monomer-monomer and monomer-solvent interactions, and by temperature~\cite{Bhattacharjee_2013, Care2014, care2015, Lesage_2019}.   

This polymer theory perspective might be revisited in the context of a better understanding of LLPS around chromatin. 
In the classical polymer literature, the folding states of the polymer, coil or globule, are related to the solvent quality : polymers expand in \textit{good} solvents, or condense in \textit{poor} solvents~\cite{DeGennes1975}. 
The nucleoplasm is a particularly complex solvent for chromatin, due to the variety and high density of macromolecules. LLPS separates several mixtures of proteins and nucleic acids that may be considered as good or poor solvents at mesoscales, potentially leading to a variety of rich phenomena. 

In non biological contexts, the behavior of polymers in much simpler solvent mixtures has been extensively studied by chemists. Non trivial effects have been described, which are fully ignored by the biophysical scientific community. 
In particular, co-non-solvency describes the collapse of a polymer chain in a good solvent when it is mixed with a second good solvent. The symmetric effect, co-solvency, refers to the swelling of a polymer chain in poor solvent mixtures~\cite{Mukherji_2014, Mukherji_2018, Bharadwaj_2019, Zhang_2024}. 
Such phenomena have been observed in the absence of LLPS-driven mesoscale droplets, since phase separating droplets in equilibrium conditions are growing under the effect of Oswald ripening and coalescence. The resulting macroscopic phases, ultimately separated by gravity, have a typical size that is much larger than the gyration radius of a polymer. Therefore, the relevance of co-solvency/co-non-solvency for chromatin systems \textit{in vivo} is yet to be characterized, with a non trivial role of mesoscale droplets. 

Regarding the role of \emph{chromatin-binding proteins} on chromatin folding, some authors have suggested that the formation and condensation of chromosome domains might be explained by a local phase separation mechanism driven by attractive DNA-protein interactions, named bridging-induced phase separation (BIPS)~\cite{Ryu_2021,brackley2013nonspecific}. 
BIPS builds on the common-sense idea that some proteins can act as a glue to condense chromatin. 
Within this framework, a protein with multivalent DNA interactions bridges two distinct regions of DNA. The resulting 
loop formation increases the local concentration in DNA and in DNA-binding proteins, hence driving phase separation. It is important to note that in the BIPS framework, phase separation \textit{only} occurs when proteins are mixed with chromatin, which differentiates BIPS from LLPS. 
It is seldom noticed that this perspective strongly contrasts with the classical model of a polymer in a good solvent, where attractive monomer-solvent interactions lead to the extension of the polymer chain. 
While the existence of proteins that help chromatin condensation is clear, and supports BIPS, there are also many observations of the swelling of chromatin fragments upon the recruitment of specific DNA-binding proteins~\cite{sellou2016poly, smith2023hpf1}. A fundamental question emerges : when do DNA-binding proteins act as a glue (a \emph{binder}) and when do they act as a good solvent ?

In this context, our aim is to provide a physical model to help clarifying the role of LLPS in chromatin organization. 
More precisely, we focus on revisiting the physics of polymers in binary solvents, in conditions under which such solvents may undergo LLPS resulting in droplets, whose size is similar to that of a chromatin domain. 
We build a minimal mean-field approach, which only includes the essential ingredients to recover a coil-globule transition at thermodynamic equilibrium. In contrast with most polymer studies~\cite{DeGennes1975,grosberggiant,Victor1990,Lesage_2019}, our model does not relay on writing the free energies as function of the gyration radius or related order parameters, since it cannot be easily coupled to the free energy of a fluid undergoing LLPS (the attempts to do so rely on strong approximations~\cite{Rouches_2024}). Instead, we consider a multicomponent interacting mixture of three species, (m, S, B), hereafter named, the monomers (m), the solvent (S) and the droplet material (B). In addition, the monomers diffuse under the action of a harmonic potential oriented towards the center of the system. This model represents a polymer through its most fundamental constraint: monomers are effectively bound to the center of mass of the polymer. The governing equations are the continuity (or diffusion) equations for the probability distribution (or volume fraction field) of each species. We here focus on the equilibrium solutions of such equations, with a particular interest in the variance, defined as the second moment of the monomer probability distribution, which plays the role of the square of the radius of gyration of a polymer chain. The main qualitative results of our model are compared with Brownian Dynamics simulations of a bead-spring polymer in a demixing Lennard-Jones fluid. 

In section \ref{sec:binary}, we study the case where monomers are dispersed into a one-component solvent, S. At equilibrium, we define two distinct states: a swollen ``coil$"$ and a dense ``globule$"$ when the monomers are either dilute or dense around the center of the system, respectively. 
In section \ref{sec:ternary}, we investigate the effects of a third species, B. When B and S demix, we show that the resulting LLPS-driven droplet of B affects the size of the coil or globular state of the monomers. We define regions in the parameter space where nonmonotonicity of the variance for the monomer spatial distribution with respect to the added amount of B is found. In particular, we observe a strong co-non-solvency effect, when the monomers that mix well with S follow a distribution that expands when increasing the concentration of a second good \emph{solvent}, here B. In the same section, we support our findings by means of Brownian dynamics simulations of explicit polymer chains surrounded by a droplet-forming fluid of Lenard-Jones particles. The last section is devoted to the conclusions.

\section{\label{sec:binary} binary model: a regular solution under confinement displays a coil-globule crossover}

\subsection{\label{sec:binary_1_a} General framework}

We consider a closed system at fixed volume $\Omega$ and temperature $T$ composed of a mixture of two species: a species \emph{m} of number $N_m$ and the \textit{solvent} species, S. Hereafter, the species \emph{m} are referred to as monomers, by analogy with the physics of a polymer chain (more detailed in section \ref{sec:binary_2_2}).

In the framework of the thermodynamics of irreversible phenomena, neglecting any barycentric motion, the balance (or continuity) equation for the volume fraction of the monomers $\phi_m = N_m v/\Omega$ reads~\cite{Groot_1984}

\begin{equation}
	\frac{\partial \phi_{m}}{\partial t} (\underline{r},t)  = - \nabla \, \underline{j}^{\mathrm{dif}}_{m} (\underline{r},t),
\label{eq:eq_1}
\end{equation}
where the diffusive flux of the monomer species, $\underline{j}^{\mathrm{dif}}_{m}$, reads, \textcolor{black}{by neglecting cross-diffusion},
\begin{equation}
	\underline{j}^{\mathrm{dif}}_{m} =   - \Lambda  \nabla \frac{ \left( \mu_m - \mu_S\right)}{T}.
\label{eq:eq_2}
\end{equation}
with $\Lambda$ the diagonal Onsager coefficient of the monomer, $\mu_m$ and $ \mu_S$ the chemical potentials of the monomer and solvent species, respectively. The thermodynamic driving force of the monomer diffusion flux is the gradient of the \textit{exchange} chemical potential, $\mu =  \mu_m - \mu_S$. The balance equation for the solvent is straightforwardly deduced from Eq. (\ref{eq:eq_1}) from the conservation law, $\phi_m +\phi_S = 1$, $\forall (\underline{r},t)$.

To model the non-uniform density of the polymer chain in the system, we assume the monomers are diffusing under the action of a fixed external potential $U$, which breaks the system translational symmetry. If the mixture is a regular solution~\cite{Prigogine_1954}, $i.e.$ characterized by an excess mixing enthalpy but an ideal mixing entropy, the exchange chemical potential then reads~\cite{Cahn_1958, Berry_2018, Kirschbaum_2021,Kirschbaum_2022} (see supplementary material) 

\begin{equation}
	\begin{split}
\mu &= {w_m - w_S} + k_{\mathrm{B}}T \mathrm{ln}\frac{\phi_m}{1-\phi_m} + k_{\mathrm{B}}T  \chi (1-2\phi_m) \\
&\ -\kappa \nabla^2 \phi_m +U,
\label{eq:eq_3}
\end{split}
\end{equation}
where $\kappa = k_{\mathrm{B}}T v^{2/d} \chi$, with $v$ the volume occupied by the monomer, $\chi = \frac{z}{2 k_\mathrm{B}T} (2 \epsilon_{mS} - \epsilon_{mm} - \epsilon_{SS})$, with $z$ the coordination number and $\epsilon_{ij}$ the mean-field pairwise interaction energy between the species $i$ and $j$. $w_m$ and $w_S$ are the internal energies of the monomer and solvent species.
The interaction parameter $\chi$ quantifies the deviation from an \emph{ideal} mixture, for which $\chi =0$. 

As we proceed to show, Eq. (\ref{eq:eq_3}) encodes three essential ingredients for a coil-globule transition: a (nonhomogeneous) monomer distribution in all the parameter space (which is induced by the inclusion of the external potential $U$), excluded-volume effects (given by the logarithmic entropic contribution), 
and energetic interactions (which are the terms proportional to the interaction parameter, $\chi$).

If excluded volumes and interactions are neglected, we wish to recover an ideal (Gaussian) chain~\cite{Bhattacharjee_2013}. This requires the external potential 
$U$ to be harmonic. This potential is centered around the center-of-mass of the system, located at the center of the system, $\underline{r}_c$. It reads

\begin{equation}
U \left(\underline{r}-\underline{r}_c\right) = - k_{\mathrm{B}}T \mathrm{ln} P \left(\underline{r}-\underline{r}_c\right),
\label{eq:eq_4}
\end{equation}
where $P \left(\underline{r}-\underline{r}_c\right)$ is a Gaussian distribution of the form

\begin{equation}
	P \left(\underline{r}-\underline{r}_c\right) = \frac{1}{{M}} \mathrm{exp}  \left(-\alpha \lvert  \underline{r}-\underline{r}_c  \lvert^2   \right),
\label{eq:eq_5}
\end{equation}
where ${M}$ is a normalization constant, $ {M} = (\pi/\alpha)^{d/2}$ and $\alpha = (d/2) /(2 N_{m}l^2)$, and $l = v^{1/d}$ is the nearest neighbor distance between two particles. 

From Eqs. \ref{eq:eq_1}-\ref{eq:eq_5}, the binary model reads 

\begin{equation}
		\frac{\partial \phi_{m}}{\partial t} (\underline{r},t) = \frac{D_{m}}{k_\mathrm{B} T} \nabla \, \left[ \phi_{m} \nabla \mu \right],  \\
\label{eq:eq_6}
\end{equation}
where $D_{m} = \Lambda k_{\mathrm{B}}/\phi_{m}$ is the diffusion coefficient.
 
\subsection{\label{sec:binary_2_1} Dimensionless form of the binary model and numerical implementation}
 
To analyze the equilibrium solutions of the binary model (Eq. (\ref{eq:eq_6})), we rewrite it in a dimensionless form. To do so, we define a characteristic length scale, $L_c =: l = \sqrt{D_mt_c}$, where $l = v^{1/d}$, with $v$ the molecular volume, and a characteristic time scale, $t_c$. The dimensionless space and time variables then read $x' = x/L_c$ and $t'=t/t_c$, and Eq. (\ref{eq:eq_6}) becomes (dropping all the primes)

\begin{equation}
	\frac{\partial \phi_m}{\partial t} (\underline{r},t) = \nabla \, \left[ \phi_m \nabla \mu _{U=0} + \phi_m \nabla U \right], 
\label{eq:eq_12}
\end{equation}
where we have separated the contribution for the potential in the expression of the exchange chemical potential, $\mu = \mu _{U=0} + U$, which is now dimensionless and thus given by Eq. (\ref{eq:eq_3}) with $k_{\mathrm{B}}T = 1$, and where,
\begin{equation}
\nabla U =  \frac{d \ l^{d-2}}{2 N_m} \left( \underline{r}-\underline{r}_c \right).
\label{eq:eq_13}
\end{equation}

Eqs. \ref{eq:eq_12}-\ref{eq:eq_13} are analyzed analytically and solved numerically using the finite-element based COMSOL Multiphysics software on a two-dimensional domain of size $L$ \cite{COMSOL_2021}. 

The initial condition for the volume fraction of the monomer reads $\phi_m (\underline{r}) = \bar{\phi}_{m}$, where $\bar{\phi}_{m}$ is the surface average of $\phi_{m}$. Since $N_m = \bar{\phi}_{m} L^2$ by definition, Eq. (\ref{eq:eq_13}) becomes 

\begin{equation}
\nabla U =  \frac{1}{\bar{\phi}_{m} L^2} \left( \underline{r}-\underline{r}_c \right).
\label{eq:eq_13b}
\end{equation}

As classically used in diffusion models with phase instabilities, we could also use an initial condition of the form $\phi_m (\underline{r}) = \bar{\phi}_{m}+\delta \phi_m$, where $\delta \phi_m$ is sampled from a uniform (or normal) distribution. Such a perturbation $\delta \phi_m (\underline{r})$ is however not required in the present case, since the presence of the harmonic potential already breaks the translational symmetry in the binodal/spinodal regions where phase separation emerges. 

The boundary condition as applied to each solid boundary of the domain is a wetted wall condition, which is suitable for solid walls in contact with a fluid interface. It sets the normal component of the (pure) diffusive flux of the monomer to zero $\underline{n} \, \underline{j}^{\mathrm{dif}}_{m, U = 0} \rvert_{\mathrm{wall}} = 0$, where $\underline{n}$ is the unit normal vector to the wall and $ \underline{j}^{\mathrm{dif}}_{m, U = 0} = D_m \, \phi \nabla \mu _{U=0}/k_\mathrm{B} T $, and it adds a wettability condition, $\underline{n} \, \nabla{\phi}_m\rvert_{\mathrm{wall}} = \mathrm{cos} (\theta_w) \lvert \nabla{\phi}_m \rvert $, where $\theta_w$ is the contact angle. For a partially wetted solid wall in contact with the fluid, we put $\theta_w = \pi/2$ (rad). 
This type of boundary conditions is found to be more stable numerically than others, such as periodic boundary conditions, using the COMSOL Multiphysics software~\cite{COMSOL_2021}. 

In order the check the reliability of our numerical procedure, in addition to usual convergence tests, we have established some expected properties of the solutions. From Eqs. \ref{eq:eq_12}-\ref{eq:eq_13b}, the equilibrium solutions are obtained by the balance of the two forces, $\nabla \mu_{U=0} = -\nabla U$, acting on the monomer field. At equilibrium, $\forall \underline{r} = (x,z)$, we find that $\phi_m (\underline r, t)$ must satisfy the general relation (see supplementary material)

\begin{equation}
	\begin{split}
	&\frac{\phi/(1-\phi)}{\langle \phi/(1-\phi) \rangle_{x,z}} (\underline{r}) = \\
	&\ \frac{ \mathrm{exp}  \left(-\alpha \lvert  \underline{r}-\underline{r}_c \lvert^2\right)  \times \mathrm{exp}  \left(2\phi\chi + \chi \nabla^2 \phi \right)}{\langle \mathrm{exp}  \left(-\alpha \lvert  \underline{r}-\underline{r}_c \lvert^2\right)  \times \mathrm{exp}  \left(2\phi\chi + \chi \nabla^2 \phi \right) \rangle_{x,z}},  \\
\label{eq:eq_14}
\end{split}
\end{equation}
where $\alpha = 1/(2\bar{\phi}_mL^2)$, $\phi$ denotes the equilibrium monomer volume fraction, $\phi = \phi_m (\underline{r}, t \rightarrow \infty)$, and where $\langle f \rangle_{x,z}$ is the surface average of a given function $f$ defined as $\langle f \rangle_{x,z} = \int_{0}^{L} \int_{0}^{L} f(x,z) dxdz/L^2$. 

In particular, in the infinite dilution limit ($\chi = 0$ and $\phi << 1$, $\forall \underline{r}$), $\phi$ follows a Gaussian distribution
\begin{equation}
\phi (\underline{r}) = \frac{\mathrm{exp}  \left(-\alpha \lvert  \underline{r}-\underline{r}_c \lvert^2\right)  }{2\pi},
\label{eq:eq_15}
\end{equation}
as expected (see section \ref{sec:binary_1_a}). In general, Eq. (\ref{eq:eq_14}) describes a non-Gaussian spatial distribution for the monomer density field.

\subsection{\label{sec:binary_2_2} Results : Our minimal model qualitatively reproduces the coil-globule crossover}

\begin{figure}[]						
\begin{center}
\includegraphics[width=8cm,height=8cm,keepaspectratio]{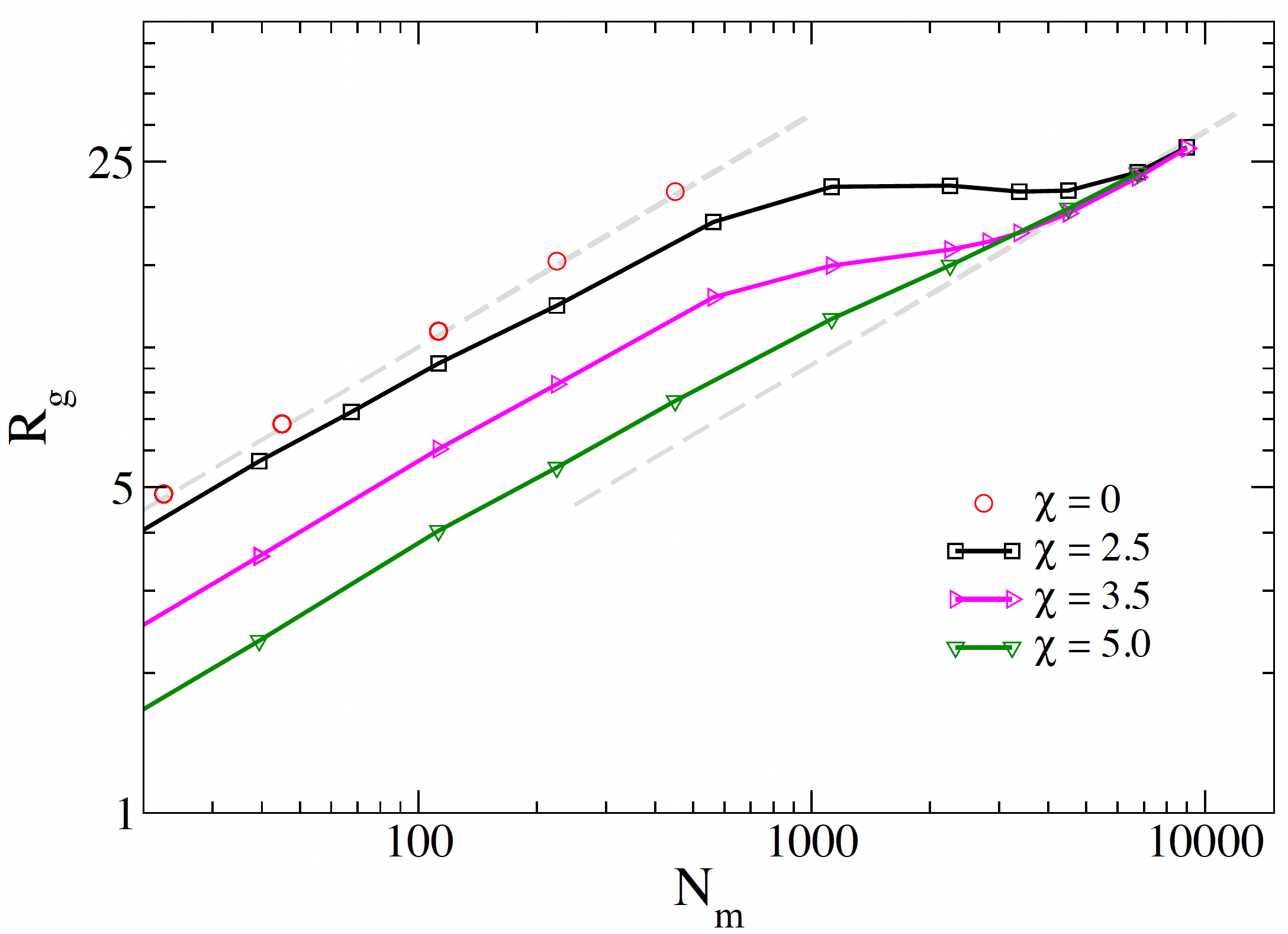}
\caption{{Radius of gyration, $R_g$, as a function of the number of monomers $N_m = \bar{\phi}_m L^2$, with $L = 150$, for different values of the interaction parameter, $\chi$ (double-logarithmic scales).} The two dashed lines correspond to the standard deviations of a Gaussian chain and of a uniform distribution of equations, $R_g = \sqrt{N_m}$ and $R_g = 0.29 \sqrt{N_m}$, respectively.} 
\label{fig:fig_1}
\end{center}
\end{figure}

To characterize the structure of the equilibrium systems, we introduce the variance of the monomer distribution function as 

\begin{equation}
	\begin{split}
	R^2_{g} = \frac{\int_{0}^{L} (x-x_c)^2 \langle \phi \rangle_{z} dx}{\int_{0}^{L} \langle \phi \rangle_{z} dx },  \\
\label{eq:eq_16}
\end{split}
\end{equation}
with $\langle \phi \rangle_{z}$ defined as

\begin{equation}
\langle \phi \rangle_{z} = \frac{1}{L} \int_{0}^{L} \phi(x,z) dz,
\label{eq:eq_16_b}
\end{equation}
The radius of gyration $R_g$ quantifies the extension of the spatial distribution of the monomers  with respect to the center-of-mass of the system, which is here fixed at $(x_c, z_c) = (L/2,L/2)$. The evolution of $R_{g}$ with the number of monomers, $N_m = \bar{\phi}_m L^2$, is plotted in Fig.~\ref{fig:fig_1} for different values of the interaction parameter $\chi$, at fixed box size $L = 150$. The latter is typically chosen to be large compared to the spatial extent of the monomer distribution ($i.e.$, $L >> R_g$), in order to prevent any qualitative effect on our results. Nevertheless, for some values of the parameters, the finite numerical value of the monomer density at boundaries leads to a small but measurable boundary effects on $R_g$. To account for such effects, we correct Eq. (19) by replacing $\langle \phi \rangle_{z}$(x) by its reduced quantity, $\langle \phi \rangle_{z} (x) - \langle \phi \rangle_{z} (x=L)$ (or equivalently, $\langle \phi \rangle_{z} (x) - \langle \phi \rangle_{z} (x=0)$).

\begin{figure}[]
\begin{center}
 \makebox[0pt]{ \includegraphics[width=8cm,height=8cm,keepaspectratio]{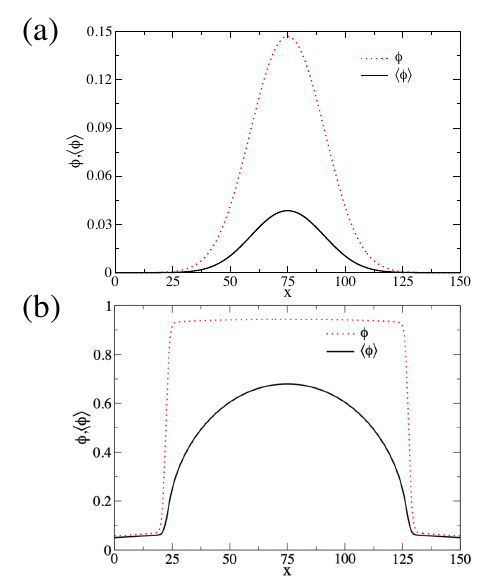}}
  \caption{{One-dimensional profile of the equilibrium monomer volume fraction $\phi$ and of the depth-averaged, $\langle \phi \rangle$, at $z=L/2$, with $L = 150$.} The cases $(a)$ and $(b)$ illustrate the two regimes of the binary model: (a) the Gaussian regime for small $\chi$ ($\bar{\phi}_m = 0.01$, $\chi = 0$). (b) the globular regime for large $\chi$ ($\bar{\phi}_m = 0.40$, $\chi = 3.0$). Equivalently, cases (a) and (b) correspond to points that lie on the top and bottom dashed lines of Figure~\ref{fig:fig_1}, respectively.}
  \label{fig:fig_2}
  \end{center}
\end{figure}

Two theoretical regimes are readily obtained analytically. 
Outside the binodal region (small $\chi$), the monomers disperse in a confined but dilute phase. This corresponds to a first regime, similar to a $coil$ polymer state.
In particular, such a coil behaves as a \textit{Gaussian} chain for an infinitely diluted mixture ($\chi = 0$ and $\phi << 1, \forall \underline{r}$). Based on Eq. (\ref{eq:eq_15}), the variance reads $R^2_g = 1/2\alpha$, which gives $R_g = \sqrt{N_m}$. In Fig.~\ref{fig:fig_1}, we find that the Gaussian regime is an excellent approximation when $\chi = 0$. The one-dimensional profile for the monomer volume fraction is shown in Fig. \ref{fig:fig_2} (a).

Inside the binodal region (large $\chi$), the monomers condense in a collapsed or globular state. For a fully condensed globule, the monomers collapse in a disk of radius $R_{circle}$ centered around the center of mass $(x_c, z_c) = (L/2, L/2)$ (see Fig.\ref{fig:fig_2} (b)).  The conservation of the total number of monomers leads to $\pi R_{\mathrm{circle}}^2 = N_m =: \bar{\phi}_m L^2$. Since the monomer distribution in the globule state follows a continuous uniform distribution of unitary height, the variance is given by $R^2_{g}=(2R_{\mathrm{circle}})^2/12$, and thus, $R_{g} = \sqrt{4 R_{\mathrm{circle}}^2/12} =: \sqrt{N_m/3\pi}$. This gives a second regime, similar to a \textit{globular} polymer state. 

In 2D systems, the coil and globule regimes are characterized by the same scaling exponent of $1/2$ with respect to $N_m$, with a different prefactor. Interestingly, our minimal model presents \emph{finite size effects} that result in a continuous change of polymer state from coil to globule as the number of monomers $N_m$ increases, for all cases when $\chi >2 $ (numerically, such a threshold is noted to be 2.1 $\pm$ 0.1 above which phase separation occurs). This transition from coil to globule with the size of the chain is usually referred to as a coil-globule \emph{crossover}. Recent analysis of super-resolution images have revealed that epigenetic domains may be well modeled by polymers close to this \emph{crossover} regime~\cite{Lesage_2019}.  The presence of this regime in our minimal model supports its relevance as a qualitative model of epigenetic domain. More precisely, a nonlinear evolution of $R_g$ $vs.$ $N_m$ marks the transition from the coil to the globule regimes as shown in Fig. \ref{fig:fig_1}. 

At fixed $\chi$, the continuous transition is characterized by an inflexion point, at a particular value for the average monomer volume fraction or equivalently, for the number of monomers. Numerically, this point is located inside the binodal region. For instance, when $\chi = 2.5$, the inflexion point for the average monomer volume fraction occurs at 0.14 $\pm$ 0.01 (or equivalently, at $N_m$ = 3150 $\pm$ 225), and when $\chi = 3.5$, it is noted at 0.09 $\pm$ 0.01 (or equivalently, at $N_m$ = 2025 $\pm$ 225). These values are smaller than those evaluated in the absence of the potential (whose analytical solution is presented in the supplementary material). 
However, we still note that phase separation can only occur when $\chi > 2.1$ $\pm$ 0.1 and thus is unchanged by the presence of the potential.
This is a clear quantitative difference with real polymers, since the critical temperature for the coil-globule transition (called the \emph{theta} temperature or $\Theta$ \emph{point}) is much higher than the corresponding critical temperature of a regular solution~\cite{Victor1990, Care2014}. Considering this limitation, all the new qualitative results obtained with our minimal model, and presented in the following sections, will be compared to state-of-the-art Brownian Dynamics simulations of systems with an explicit polymer chain.  

All in all, the binary model reproduces well some key equilibrium properties of a polymer chain, as accurately described by an interacting self-avoiding walk~\cite{Lesage_2019}. The common ingredients of this minimal model with a model of LLPS allow us to couple within the $same$ thermodynamic framework the two phase phenomena involved within the cell nucleus: the coil-globule transition and LLPS. When a third species, B, is introduced in the model, the mixture of B and S may phase separate and affects the equilibrium properties of the monomers in both the coil and globular states. We then expand the binary model to a ternary one.

\section{\label{sec:ternary} Ternary model: interplay of phase transitions}

\subsection{Extension of the binary model}

We now consider a system at constant volume and temperature filled with a mixture of three species, the monomers (m), the droplet material (B) and the solvent (S). The governing equations are the two coupled continuity equations for the monomers and the droplet material volume fractions

\begin{equation}
	\frac{\partial \phi_{m}}{\partial t} (\underline{r},t)  = - \nabla \, \underline{j}^{\mathrm{dif}}_{m} (\underline{r},t),
\label{eq:eq_17}
\end{equation}

\begin{equation}
	\frac{\partial \phi_{B}}{\partial t} (\underline{r},t)  = - \nabla \, \underline{j}^{\mathrm{dif}}_{B} (\underline{r},t),
\label{eq:eq_18}
\end{equation}
where the diffusive fluxes of the monomer and droplet material species are given by (neglecting cross-diffusion) 

\begin{equation}
	\underline{j}^{\mathrm{dif}}_{m} =   - \Lambda \nabla \frac{ \left( \mu_m - \mu_S\right)}{T},
\label{eq:eq_19}
\end{equation}

\begin{equation}
	\underline{j}^{\mathrm{dif}}_{B} =   - \Lambda^{\prime} \nabla \frac{ \left( \mu_B - \mu_S\right)}{T},
\label{eq:eq_20}
\end{equation}
with $\Lambda^{\prime}$ the diagonal Onsager coefficient of B and $\mu_B$ the related chemical potential. As in the binary model, we define the exchange chemical potentials, $\mu = \mu_m - \mu_S$, and $\mu' = \mu_B - \mu_S$. They read \cite{Kirschbaum_2021,Kirschbaum_2022} (see supplementary material) 
\begin{equation}
	\begin{split}
	&\mu = {w_m - w_S} +k_{\mathrm{B}}T \mathrm{ln}\frac{\phi_m}{1-\phi_m-\phi_B}   \\
	& \ + k_{\mathrm{B}}T (\chi_{mB} \textcolor{black}{-\chi_{BS}}) \phi_B + k_{\mathrm{B}}T \chi_{mS} (1-2 \phi_m - \phi_B) \\
	& \  + \bar{\kappa} \nabla^2 \phi_B - {\kappa_{mS}} \nabla^2 \phi_m +U,
\label{eq:eq_21}
\end{split}
\end{equation}

\begin{equation}
	\begin{split}
	&\mu^{\prime} = {w_B - w_S} +k_{\mathrm{B}}T \mathrm{ln}\frac{\phi_B}{1-\phi_m-\phi_B}  \\
	& \ + k_{\mathrm{B}}T (\chi_{mB} \textcolor{black}{-\chi_{mS}}) \phi_m + k_{\mathrm{B}}T \chi_{BS} (1-2 \phi_B - \phi_m)   \\
	& \  + \bar{\kappa} \nabla^2 \phi_m - {\kappa_{BS}} \nabla^2 \phi_B ,
\label{eq:eq_22}
\end{split}
\end{equation}
where $\bar{\kappa} = k_{\mathrm{B}}T  v^{2/d} \left( \chi_{mB}-\chi_{BS}-\chi_{mS}\right)$, $\kappa_{ij} =  k_{\mathrm{B}}T  v^{2/d} \chi_{ij}$, with $v$ the molecular volume, assumed to be the same for all the species, and $w_i$ the internal energies. The continuity equation for the solvent species naturally follows from the conservation law $\phi_m + \phi_S + \phi_B = 1, \forall (\underline{r},t)$. As an extension of the binary model, the ternary model now admits three interaction parameters ($\chi_{mB}, \chi_{BS}, \chi_{mS}$). 

From Eqs. \ref{eq:eq_17}-\ref{eq:eq_22} with $U$ given by Eq. (\ref{eq:eq_4}), the dimensionless ternary model reads

\begin{equation}
	\frac{\partial \phi_m}{\partial t} (\underline{r},t) = \nabla \, \left[ \phi_m \nabla \mu _{U=0} + \phi_m \nabla U \right], 
\label{eq:eq_23}
\end{equation}

\begin{equation}
	\frac{\partial \phi_B}{\partial t} (\underline{r},t) = \delta_B \nabla \, \left[ \phi_B \nabla \mu^{\prime}  \right], 
\label{eq:eq_24}
\end{equation}
with $\mu =: \mu _{U=0} + U$, and $\mu'$ given by Eqs. \ref{eq:eq_21}-\ref{eq:eq_22}, respectively, with $k_{\mathrm{B}}T = 1$, and where, in two dimensions,
\begin{equation}
\nabla U =  \frac{1}{ \bar{\phi}_m L^2} \left( \underline{r}-\underline{r}_c \right),
\label{eq:eq_25}
\end{equation}
and $\delta_B = D_B/D_m$, with $D_B = \Lambda' k_{\mathrm{B}}/\phi_B$, the diffusion coefficient of B. Since $\delta_B$ only affects the relaxation kinetics but not the equilibrium solutions, we simply put $\delta_B = 1$.

\subsection{Influence of the droplet material B on the polymer radius of gyration}

We follow the radius of gyration associated to such a distribution as a function of the average volume fraction of B, $\bar{\phi}_B$, for different interaction parameters $(\chi_{mB}, \chi_{BS}, \chi_{mS})$ and average monomer volume fraction, $\bar{\phi}_m$. Due to the conservation of the total number of particles and since the exchange chemical potential of B is defined with respect to the one of the solvent, adding B particles in the system corresponds to removing the same number of S particles (solvent) from the system. Hereafter, we classify the findings in two distinct situations, when the evolution of $R_g$ is observed to be monotonic and when it is found to behave nonmonotonically with respect to the average volume fraction of B.

\subsubsection{Monomotonic evolution of $R_g$ $vs.$ $\bar{\phi}_B$}

\begin{figure}[]						
\begin{center}
\includegraphics[width=8.5cm,height=8.5cm,keepaspectratio]{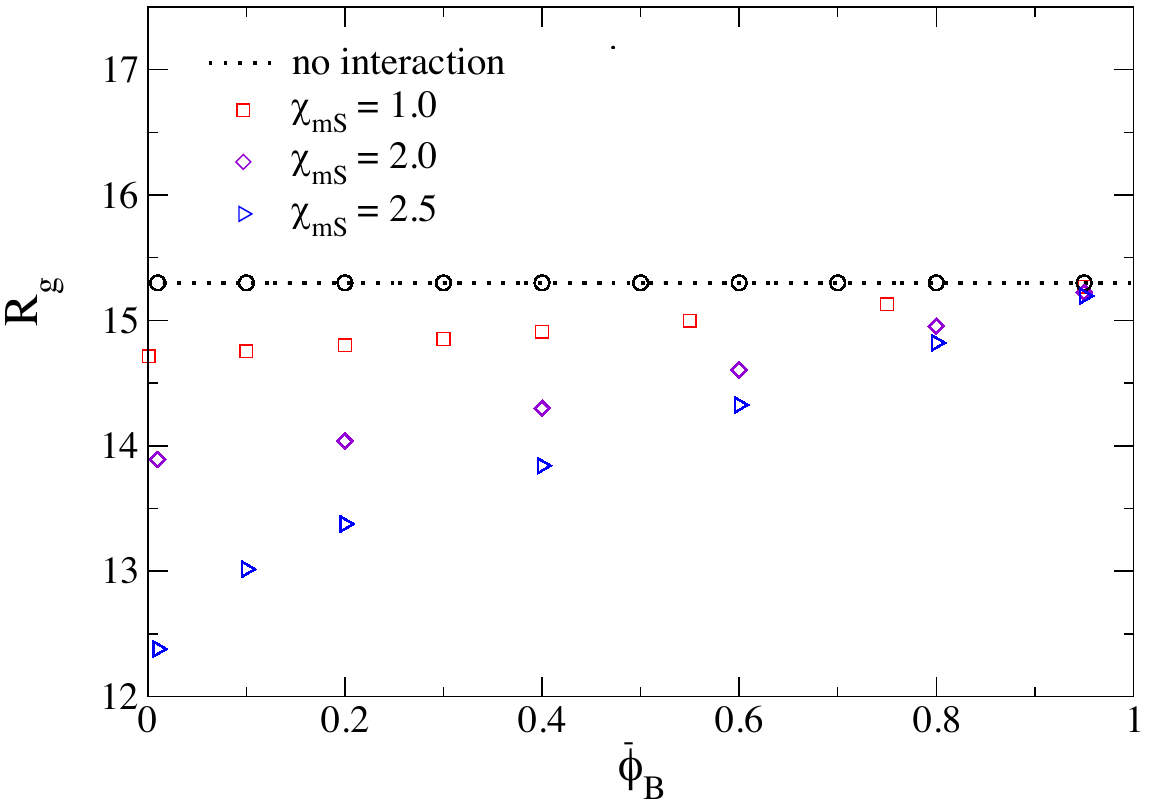}
\caption{{Radius of gyration $R_g$, as a function of the initial volume fraction of B $\bar{\phi}_{B}$, for different values of the interaction parameter $\chi_{mS}$, when  $\chi_{BS}=  \chi_{mB}$ = 0 ($L = 150$, $\bar{\phi}_m = 0.01$).}} 
\label{fig:fig_3}
\end{center}
\end{figure}

The simplest scenario corresponds to $(\chi_{mB}, \chi_{BS}, \chi_{mS}) = (0, 0, 0)$. The radius of gyration does not depend on $\bar{\phi}_B$, since the species B and S are totally equivalent for the monomers. In particular, in the range of parameters we investigated ($\bar{\phi}_m = 0.01$ and $L=150$), we find $R_g = 15.3$, $\forall  \bar{\phi}_B$ (see Fig. ~\ref{fig:fig_3}). 

\begin{figure}[]						
\begin{center}
\includegraphics[width=8.5cm,height=8.5cm,keepaspectratio]{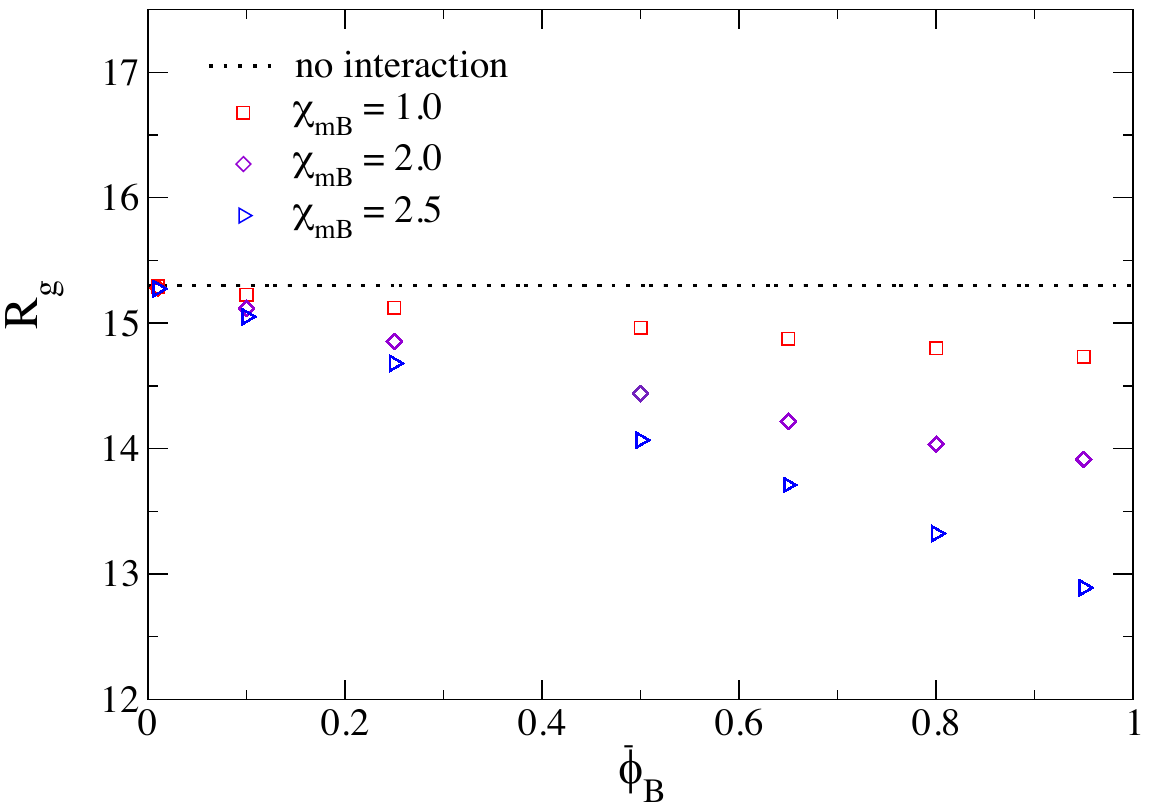}
\caption{{Radius of gyration $R_g$, as a function of the average volume fraction of B $\bar{\phi}_{B}$, for different values of the interaction parameter $\chi_{mB}$, when  $\chi_{mS}=  \chi_{BS}$ = 0 ($L = 150$, $\bar{\phi}_m = 0.01$).}}  
\label{fig:fig_4}
\end{center}
\end{figure}

Next, we investigate the case $(\chi_{mB}, \chi_{BS}, \chi_{mS}) = (0, 0, \chi_{mS} \ne 0)$. When $\chi_{mS} >0$, the net repulsion between the monomers and S results in an effective attraction between the monomers and B. Hence, the mixture (B, S) is a better solvent (for m) than the pure S fluid. Thus, by replacing B by S, the polymer expands in a better solvent. The expansion (or contraction) of the polymer when $\chi_{mS} >0$ (or $\chi_{mS} <0$) is verified $\forall (\bar{\phi}_m, \bar{\phi}_B)$ and $\chi_{mS}$ (see Fig. ~\ref{fig:fig_3}).

In the opposite case, we consider the set, $(\chi_{mB}, \chi_{BS}, \chi_{mS}) = (\chi_{mB} \ne 0, 0, 0)$. When $\chi_{mB} > 0$, the pure S fluid is a better solvent than the mixture (B, S). Hence, the polymer contracts when increasing $\bar{\phi}_B$ ($R_g$ decreases, as seen in Fig. ~\ref{fig:fig_3}). The contraction (or expansion) of the polymer when $\chi_{mS} < 0$ (or $\chi_{mS} > 0$) is verified $\forall (\bar{\phi}_m, \bar{\phi}_B)$ and $\chi_{mB}$.

\subsubsection{Nonmonotonic evolution of $R_g$ $vs.$ $\bar{\phi}_B$ : mesoscale co-non-solvency effect}

We now analyze scenarios in which the enthalpic contributions favor phase separation of the B/S mixture, when  $\chi_{BS} > 0$.

\begin{figure}[]						
\begin{center}
\includegraphics[width=8.5cm,height=8.5cm,keepaspectratio]{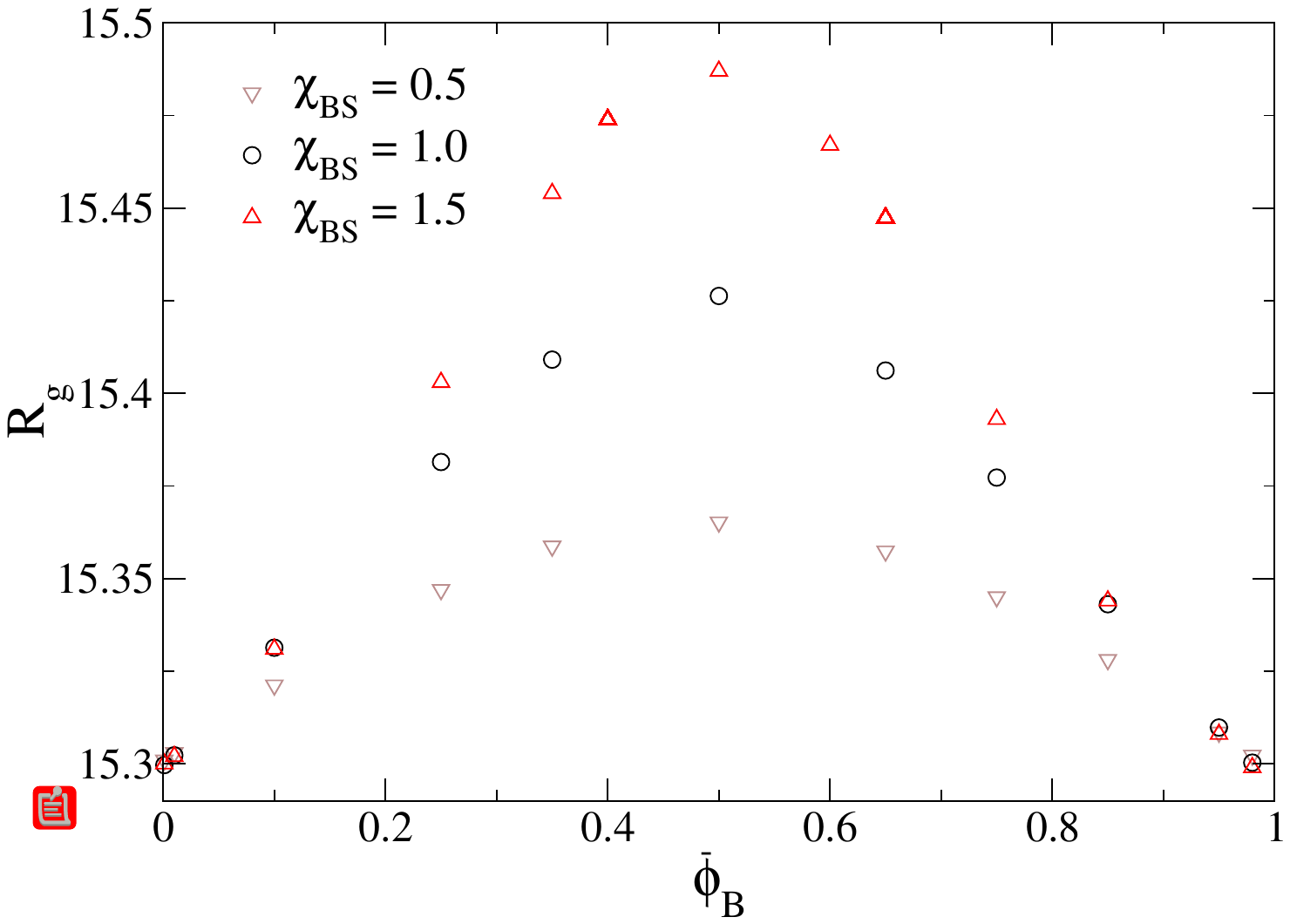}
\caption{{Radius of gyration $R_g$, as a function of the average volume fraction of B $\bar{\phi}_{B}$, for different values of the interaction parameter $\chi_{BS}$, when  $\chi_{mS}=  \chi_{mB}$ = 0  ($L = 150$, $\bar{\phi}_m = 0.01$).}}  
\label{fig:fig_5}
\end{center}
\end{figure}

To highlight the effects of $\chi_{BS}$, let us consider the set  $(\chi_{mB} = 0, \chi_{BS} > 0, \chi_{mS} = 0)$, where B and S are enthalpically and entropically equivalent for the monomers. From the conservation law $\bar{\phi}_m + \bar{\phi}_S + \bar{\phi}_B =1$, the radius of gyration changes monotonic properties with respect to $\bar{\phi_B}$ when $\bar{\phi}_B = \bar{\phi}_S$, $i.e.$, when $\bar{\phi}_B = (1-\bar{\phi}_m)/2$ (see Fig. \ref{fig:fig_5}). This result is expected as it arises from the \emph{mirror} symmetry of the ternary model when $\chi_{mB} = \chi_{mS}$. 

\begin{figure}[]						
\begin{center}
\includegraphics[width=9cm,height=9cm,keepaspectratio]{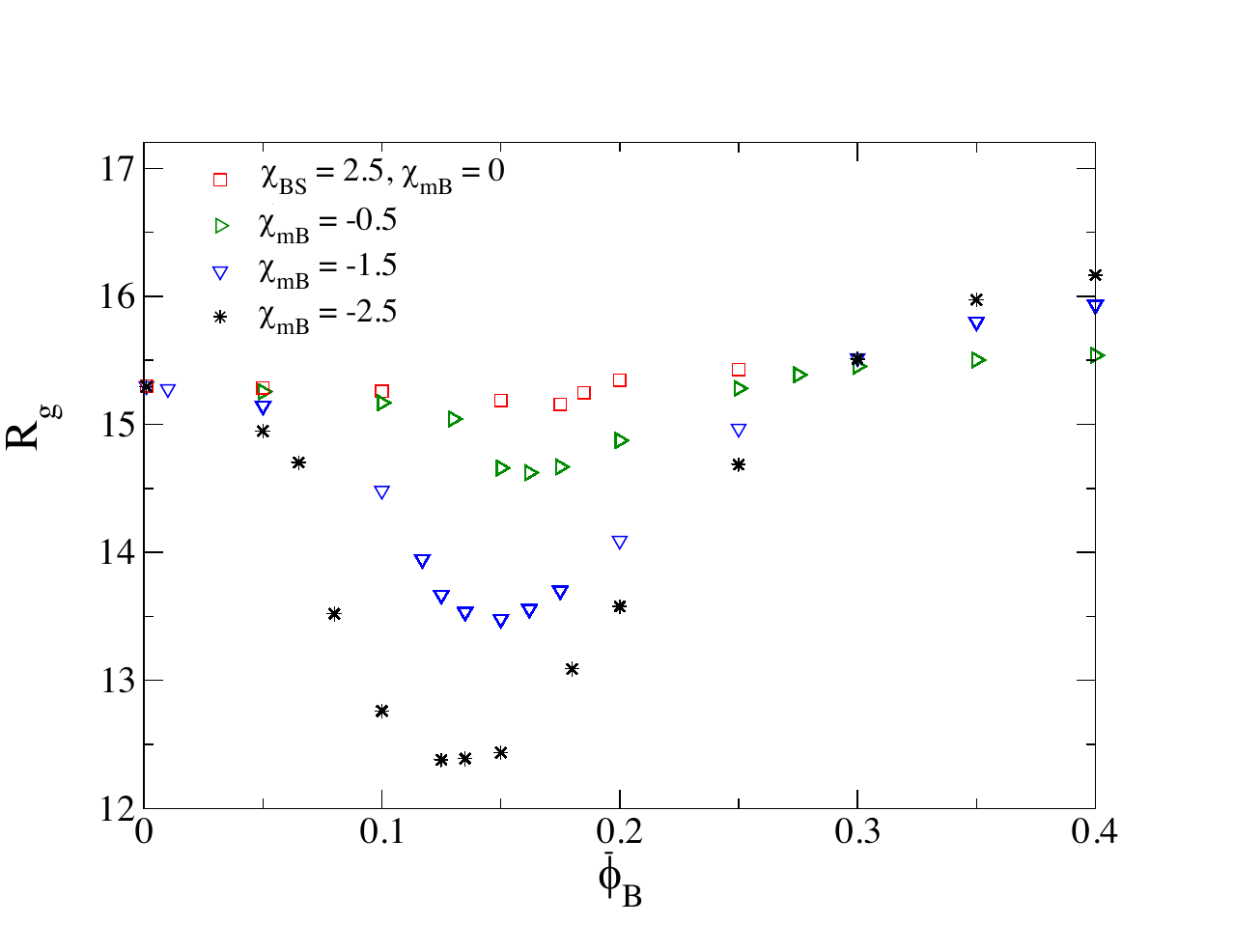}
\caption{Radius of gyration, $R_g$, as a function of the average volume fraction of B $\bar{\phi}_{B}$, for different values of the interaction parameter $\chi_{mB} < 0$, with $\chi_{BS} = 2.5$ and $\chi_{mS} = 0$ ($L = 150$, $\bar{\phi}_m = 0.01$).}  
\label{fig:fig_6}
\end{center}
\end{figure}

We then investigate values of the interaction parameters that are more likely to correspond to a droplet located on the polymer. More precisely, for values of $\chi_{BS}$ above the critical value for phase separation in absence of polymer, we explore negative values of $\chi_{mB}$, which correspond to B particles that are attracted by the polymer. 
When $\chi_{BS}$ exceeds about 2 (numerically, when $\chi_{BS} > 2.2 \pm 0.1$), we find nonmonotonicity even below the symmetry point, when $\bar{\phi}_B < (1-\bar{\phi}_m)/2$ (see Fig. \ref{fig:fig_6}). Its is characterized by the presence of a minimum of the gyration radius at a specific value of B density, $\bar{\phi}_B = \bar{\phi}^{min}_B$. 
At low $\bar{\phi}_B$, the radius of gyration is a decreasing function of $\bar{\phi}_B$. B particles induce a collapse of the polymer. A second regime appear for $\bar{\phi}_B > \bar{\phi}^{min}_B$, corresponding to an expansion of the polymer state as B particles are added. The amplitude of the size variation of the polymer can be modulated by varying $\chi_{mb}$ : this amplitude increases as the attraction between the droplet material B and monomers increases ($\chi_{mb}$ becomes more negative). 

This nonmonotonic behavior is an original form of the \emph{co-non-solvency} effect. This phenomenon has been observed in non-biological systems, when 
a polymer in a good solvent (S) is mixed with an increasing concentration (or volume fraction) of a second good solvent (B) \cite{schild1991cononsolvency, Mukherji_2014, Mukherji_2018, Bharadwaj_2019, Zhang_2024}. The situations we are investigating here are nevertheless different from the ternary systems that are described in the chemistry literature. 
In chemical systems \emph{in vitro}, at mesoscales, the polymer is dissolved in a unique phase, which is a mixture of both solvents. The typical experimental setups aim at controlling the composition of both solvents in this phase, and quantifying its influence on polymer size. In the context of our model of biocondensates, changing the quantity of B particles in the finite size system in LLPS conditions modifies the droplet size. This leads to a \emph{mesoscale co-non-solvency effect} that may be related not only on phase composition, but also on droplet size at mesoscales. Let us recall here that our fixed volume systems artificially constrain the size of the two phases. In real biological systems, the control of droplet size is achieved through non-equilibrium mechanisms~\cite{Zwicker2022}.  

\begin{figure}[]						
\begin{center}
\includegraphics[width=9.5cm,height=9.5cm,keepaspectratio]{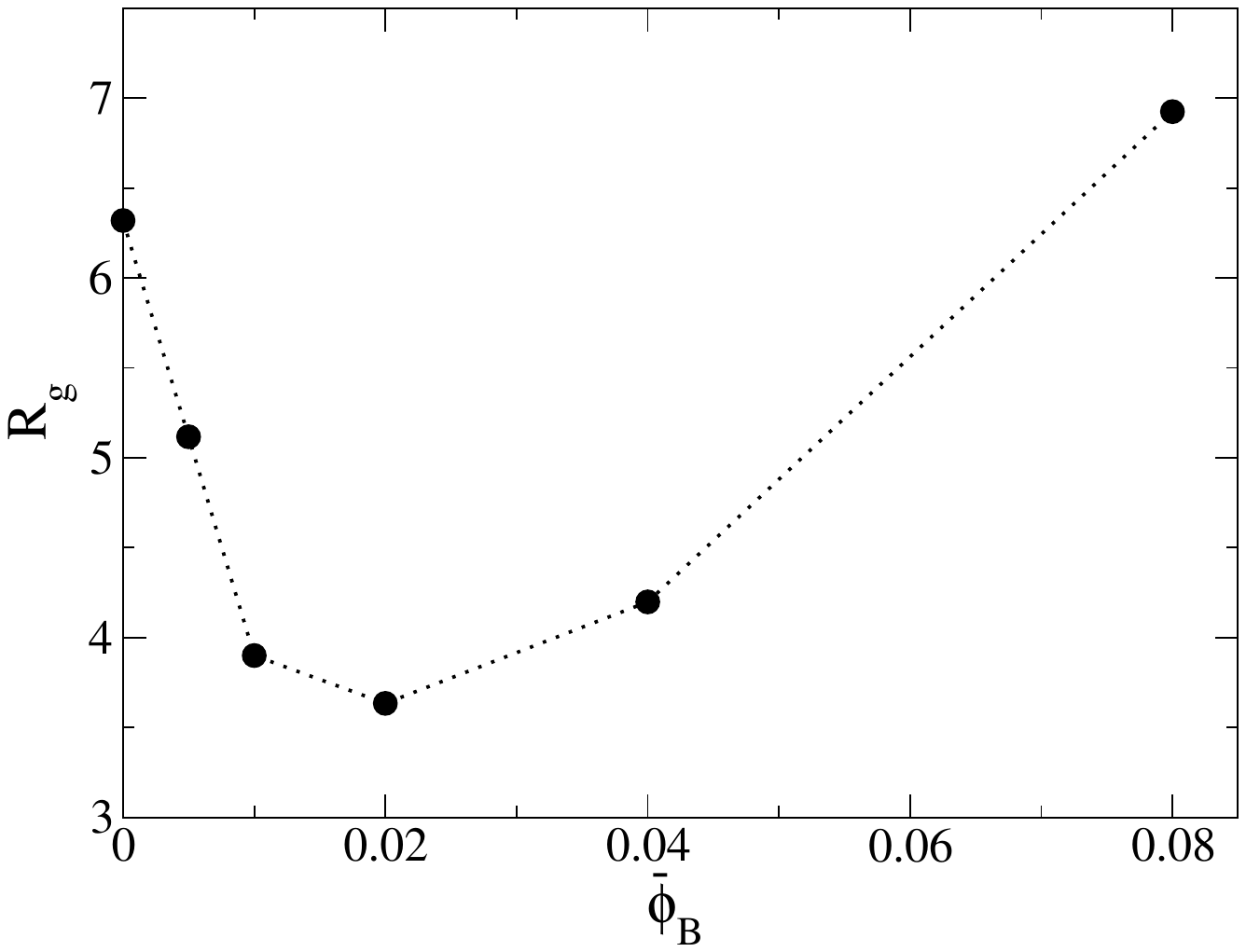}
\caption{Brownian Dynamics : Radius of gyration, $R_g$, as a function of the initial volume fraction of B, $\bar{\phi}_{B}$. The Lennard-Jones interaction potentials have energy parameters $\varepsilon_{BB} = 1$,  $\varepsilon_{mm} = 0.2$ and $\varepsilon_{mB} = 1$. The chain contains $N_m = 100$ monomers.} 
\label{fig:fig_6b}
\end{center}
\end{figure}

While these results highlight new properties of polymer systems, our minimal model does not include important ingredients that govern the specific entropic properties of polymers. The monomers are described by a regular solution model where the solutes are confined by a harmonic potential. It contains the main entropic constraint, which is the requirement that monomers stay close to each other. Nevertheless, the specific configurational entropy related to inter-monomer bounds is not well described.  
Therefore, we have checked whether the original properties of our model can be reproduced by an explicit 3D bead-spring polymer model with a full description of monomer-monomer interaction potentials by Brownian Dynamics (BD) simulations. These particle based simulations account explicitely for monomer beads of the polymer and for B solute particles, and describe the solvent only implicitely through a random term in the overdamped Langevin equation of motion. For the polymer chain we take a simple and usual model~\cite{dunweg1993molecular}: bound monomers interact by a combination of the FENE potential~\cite{wedgewood1991finitely} and the short-range repulsive Weeks-Chandler-Andersen (WCA) interaction potential~\cite{Weeks1971}. Unbound monomers and B particles interact through an attractive Lennard-Jones (LJ) potential. Monomers and B particle have the same diameter. The number of monomers is varied, as well as the volume fraction of B particles, and the LJ energy depth for interactions between B, between distant monomers, and between monomers and B, in order to investigate several regimes. All parameters and simulation details are given in the supplementary material.

In Fig. \ref{fig:fig_6b}, results from BD simulations are shown for a chain of 100 monomers in the presence of B solutes, in a case where B solutes form liquid droplets in a pure B fluid. 
 The nonmonotonic behavior as a function of the volume fraction of B particles is recovered, confirming the findings of our minimal model.

\begin{figure*}[ht!]
  \begin{tabular*}{\textwidth}{ccc}
   (a)\includegraphics[width=0.28\textwidth]{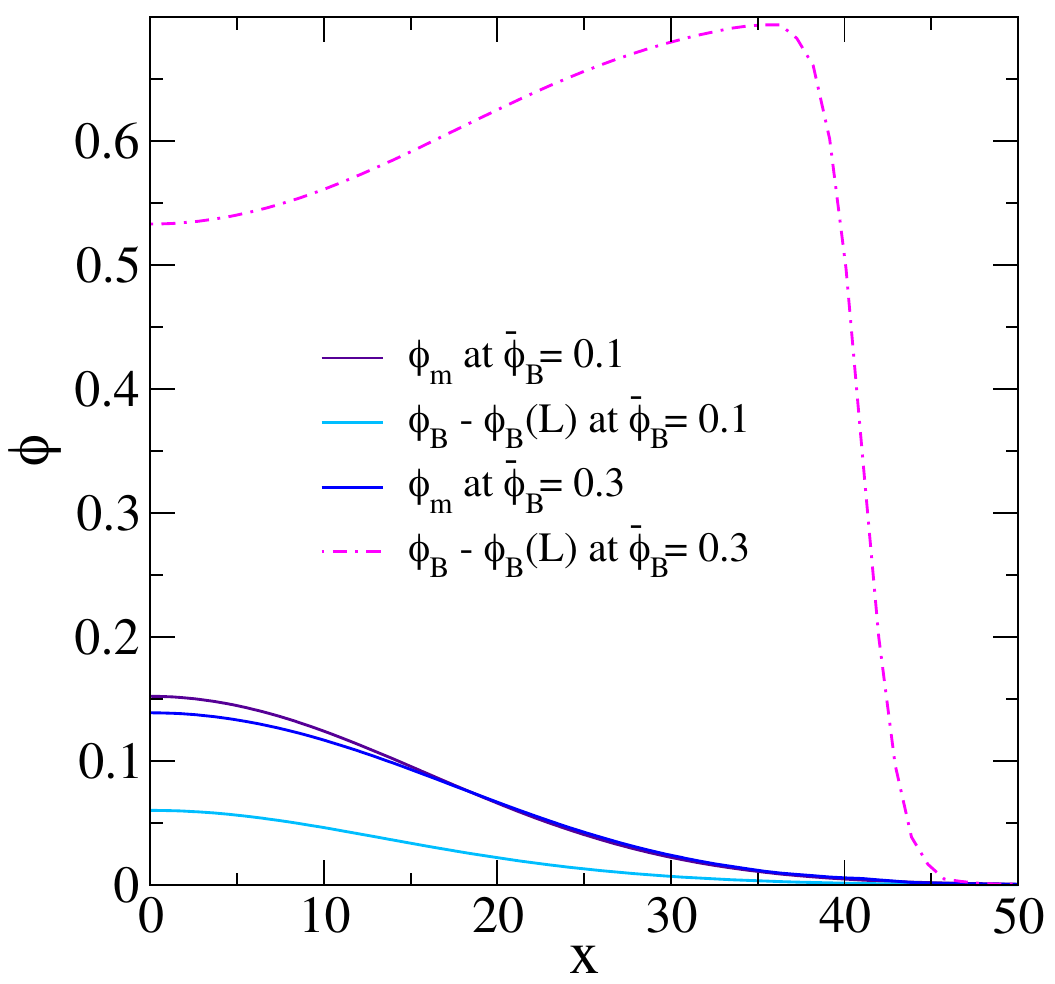} &
    (b)\includegraphics[width=0.28\linewidth]{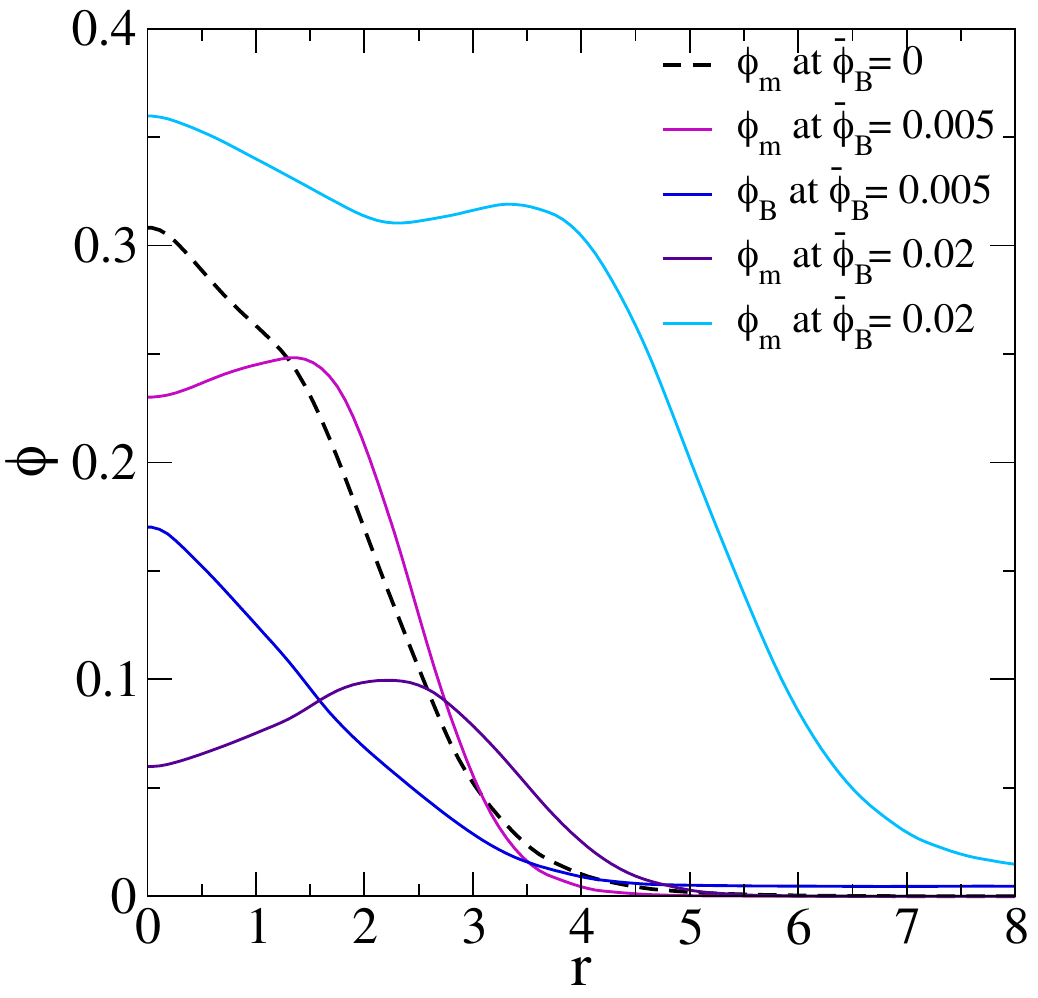} &
    (c)\includegraphics[width=0.37\linewidth]{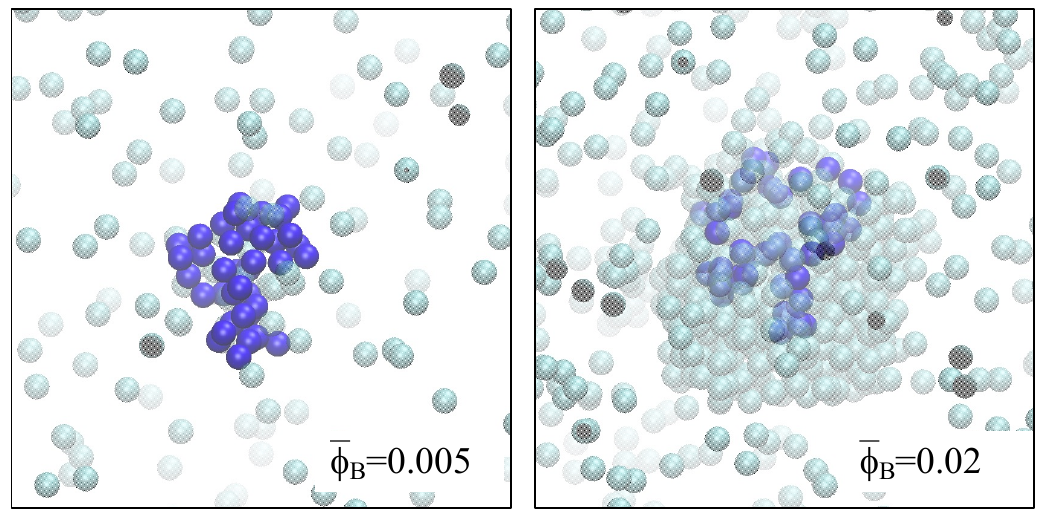} 
          \end{tabular*}
  \caption{{(a) Minimal model : Volume fraction of  droplet material, $\phi_B$, and of the monomers, $\phi_m$,   as a function of the distance $x$ to the center of the system at,  for $\bar{\phi}_B = 0.10$ and  $\bar{\phi}_B = 0.30$, when $(\chi_{mB}, \chi_{BS}, \chi_{mS}) = (2.5, -0.50, 0)$, $\bar{\phi}_m = 0.01$ with fixed $L = 150$. To facilitate the comparison with the monomers profile, the profile for the volume fraction of B is decreased by the one at the system boundary, $\phi_B$ $(x=L, z=L/2)$ (noted $\phi_B (L)$).  
 (b) Volume fraction of monomers ${\phi}_m$ and of B particles ${\phi}_B$ as functions to the distance to the center-of-mass of the polymer, for a chain of 40 monomers  at $\bar{\phi}_m = 0.0008$ obtained by Brownian Dynamics simulations for two volume fractions in B particles, $\bar{\phi}_B = 0.005$ and $\bar{\phi}_B = 0.02$, respectively smaller and larger than the minimum of Fig.~\ref{fig:fig_8b}.  The parameters of LJ interactions are $\varepsilon_{BB} = 1.2$,  $\varepsilon_{mm} = 0.6$ and $\varepsilon_{mB} = 1.2$. (c) Typical configurations of the system simulated by Brownian Dynamics at equilibrium (monomer are in dark blue, B particles in transparent cyan), at $\bar{\phi}_B = 0.005$ and $\bar{\phi}_B = 0.02$.}}
  \label{fig:fig_7}
\end{figure*}

For a better understanding of the  role of B particles, we characterized the structure of the systems for our minimal model and for the one simulated by Brownian Dynamics. Fig. \ref{fig:fig_7} (a) shows the volume fraction fields for both monomers and droplet material obtained by the minimal model, as functions of the distance to the center of the simulation box (which is also the center-of-mass of the polymer), in a case where (1) LLPS within the B/S mixture may occur, for $\chi_{BS} > 2.2 \pm 0.1$, (2) B and monomers attract each other.  
 
At low density in B particles, when $\bar{\phi}_B$ increases, most of the B particles added to the system  
tend to accumulate close to the center-of-mass of the polymer ($\underline{r}_c = (L/2,L/2)$) to avoid contacts with the S species, or equivalently, to enhance contacts with the monomers. Hence, the maximum density of B particles is located at the center of the system (see Fig. \ref{fig:fig_7} (a)). In this regime, B particles are less concentrated than monomers in the polymer region, \emph{i.e.} at distances from the center-of-mass of the polymer lower than (or similar to) the radius of gyration of the polymer. In other words, B is always surrounded by S or m particles in large excess. The m/S mixtures is the solvent of the B particles, which is enriched in monomers in the polymer region. By increasing $\bar{\phi}_B$, the replacement of the S particles by \emph{binder} B particles at the center-of-mass of the polymer drives a collapse of the polymer chain and the radius of gyration decreases. 

As $\bar{\phi}_{B}$ exceeds $\bar{\phi}^{min}_B$, the species B starts to accumulate around the polymer chain. With the minimal model, the distribution of B shows a depletion zone at the center of the system, and a global maximum at a distance from the center (see Fig. \ref{fig:fig_7} (a)). The increase of the volume fraction of B away from the center-of-mass of the polymer chain is interestingly correlated with the expansion of the polymer. This strengthens the idea that the B droplet acts as a mesoscale volume of good solvent located around the polymer. The existence of this regime is directly related to the content of B particles in the system, which is a key controlling parameter regarding the role of polymer-binding molecules on polymer folding state. In a biological context, this suggests that the same chromatin-binding protein may act either as a condensing or an expanding agent depending on its local concentration. 

The results obtained by Brownian Dynamics corroborate this picture. The concentration of both m and B particles around the center of mass of the polymer in the collapse and in the expansion regimes are shown on Fig.  \ref{fig:fig_7}(b). When B act as binders, the chain is slightly less extended in the presence of a small amount of B particles (at $\bar\phi_B=0.005$) compared to the situation in a pure solvent. The B particles are more concentrated close to the center of mass, where ther are surrounded by more concentrated monomers. The expands at high volume fraction of B (at $\bar\phi_B=0.020$), in correlation with the change of role of B particles from \emph{binders} to \emph{good solvent}: B concentration becomes much higer than monomer concentration in the polymer region. The snapshots on Fig.  \ref{fig:fig_7}(c) are taken from the simulations. They allow to visualize how the monomers behave like a solvent for B particles at small $\bar\phi_B$, whereas on the contrary, a droplet of B particles formed around the polymer at high $\bar\phi_B$ now plays the role of a solvent for the polymer chain. 

\subsubsection{Influence of finite polymer size in the presence of a mesoscale phase separation}
 
 Finite-size effects in polymers lead to significant deviations from universal scaling laws~\cite{DeGennes1975,Victor1990,care2015,Lesage_2019}.
 For instance, the coil-globule crossover is characterized by a strong influence of the size of the polymer (number of monomers). In solvent mixtures, in the presence of a biologically relevant mesoscale phase separation of distinct liquids, we have characterized in the previous section non trivial solvent effects. In this last part, we investigate how the finite-size polymer effects may couple to mesoscale solvent effects, and may give rise to qualitatively different folding behaviors for systems with similar interactions but distinct polymer sizes. 

We study systems 
for which (1) B may act as a good solvent for the polymer ($\chi_{mB} < 0$), (2) there is a coil-globule crossover of the polymer in absence of B particles ($\chi_{mS} > 2$), and (3) B particle phase separate in the solvent S, in absence of monomers ($\chi_{BS} > 2$). For such systems, at fixed $\chi_{mS}$, small polymers (low value of $\bar{\phi}_m$) behave as coils, while large polymers (high value of $\bar{\phi}_m$) behave as globules, as already pointed out in the analysis of Fig. \ref{fig:fig_1}. 
 
\begin{figure}[]						
\begin{center}
\includegraphics[width=8.5cm,height=8.5cm,keepaspectratio]{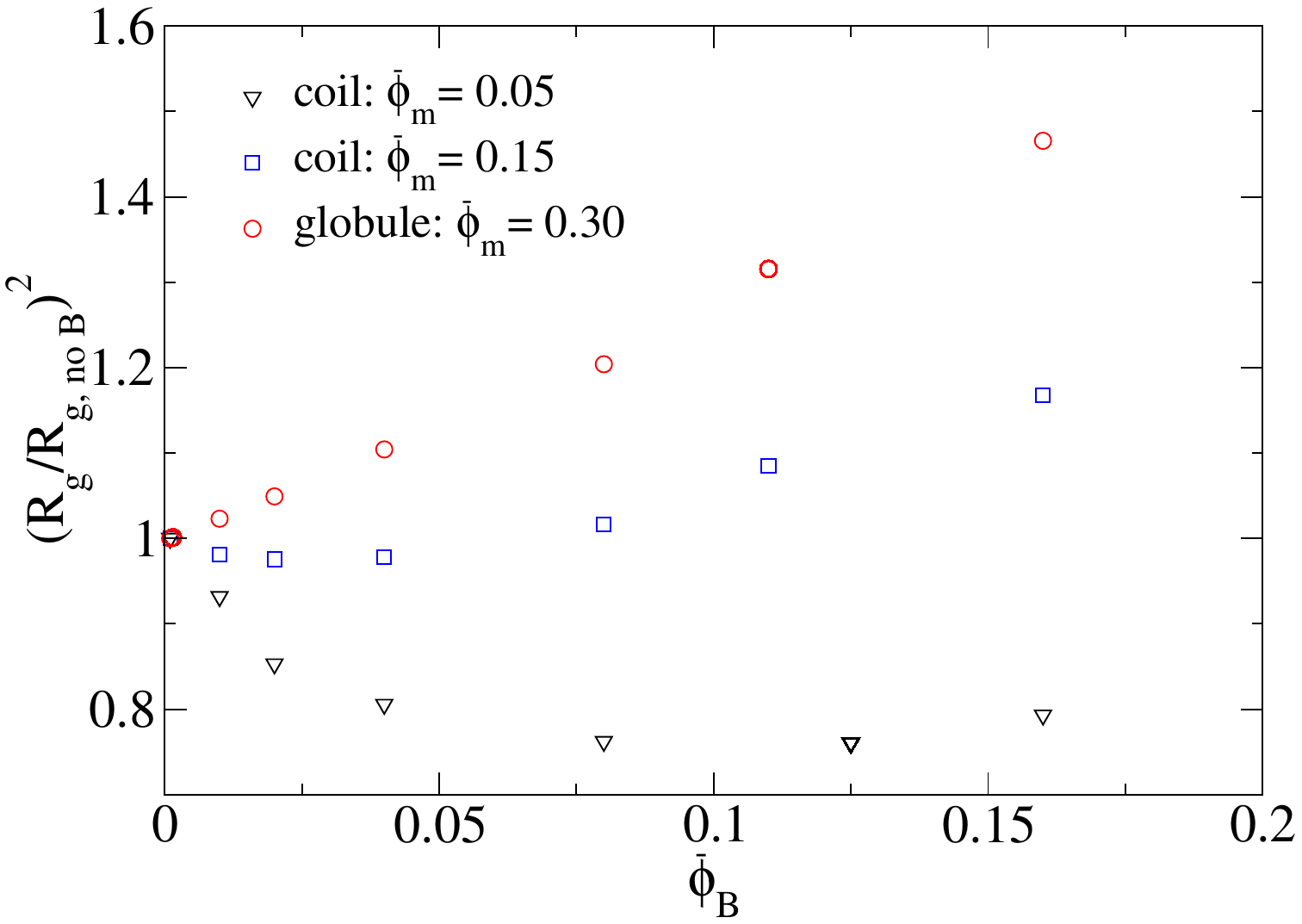}
\caption{Square of the ratio of the radius of gyration, $R_g$, with a reference value calculated in absence of droplet material, $R_{g, \mathrm{no \ B}} =: R_{g, \bar{\phi}_{B}=0}$, as a function of the average volume fraction of B, $\bar{\phi}_{B}$, when $(\chi_{mB},\chi_{BS},\chi_{mS})= (-0.50,2.5,2.5)$, for different values of the average monomer volume fraction, $\bar{\phi}_{m}$.} 
\label{fig:fig_8}
\end{center}
\end{figure}

 The evolution of the radius of gyration of the polymer with the amount of droplet material (B) is presented in Fig. \ref{fig:fig_8}, for three polymer sizes. 
 We find that the resulting curves qualitatively depend on polymer size. For the larger polymer ($\bar{\phi}_m = 0.30$), in the absence of B particles, the results from Fig. \ref{fig:fig_1} show that we are in the globular region of polymer states. In such case,  
  the radius of gyration is a monotonic increasing function of $\bar{\phi}_{B}$. The polymer expands as the droplet of good B solvent grows at its vicinity. 
  The two other systems ($\bar{\phi}_m = 0.05$ and $\bar{\phi}_m = 0.15$) correspond to coil states in absence of B. In these systems, the influence of $\bar{\phi}_{B}$ on polymer size is nonmonotonic. For the smallest polymer, when B particles are added to the system, there is a significant condensation of the polymer. At larger $\bar{\phi}_{B}$, our results show a change of behavior, with a slight expansion of the polymer when $\bar{\phi}_{B}$ increases. 
  For the intermediate polymer size ($\bar{\phi}_m = 0.15$), the condensation regime is less pronounced, while the expansion is more significant. 
  These tendencies are qualitatively similar to the mesoscale co-non-solvency effect described in the previous section. However, for the cases investigated in this section, S is a poor solvent and B is a good solvent. Finite size effects make possible the existence of a co-non-solvency for solvents of opposing qualities. As far as chromatin is concerned, finite size effects have been shown to be important~\cite{care2015,Lesage_2019}. Our new results suggest that these effects may be the consequence of a qualitatively different influence of mesoscale biocondensates on small vs. large chromatin domains. The exact same proteins may expand a large chromatin domain and condense a small chromatin domain, while the effective interactions between the macromolecules at play are identical. 

 \begin{figure}[]						
\begin{center}
\includegraphics[width=10cm,height=10cm,keepaspectratio]{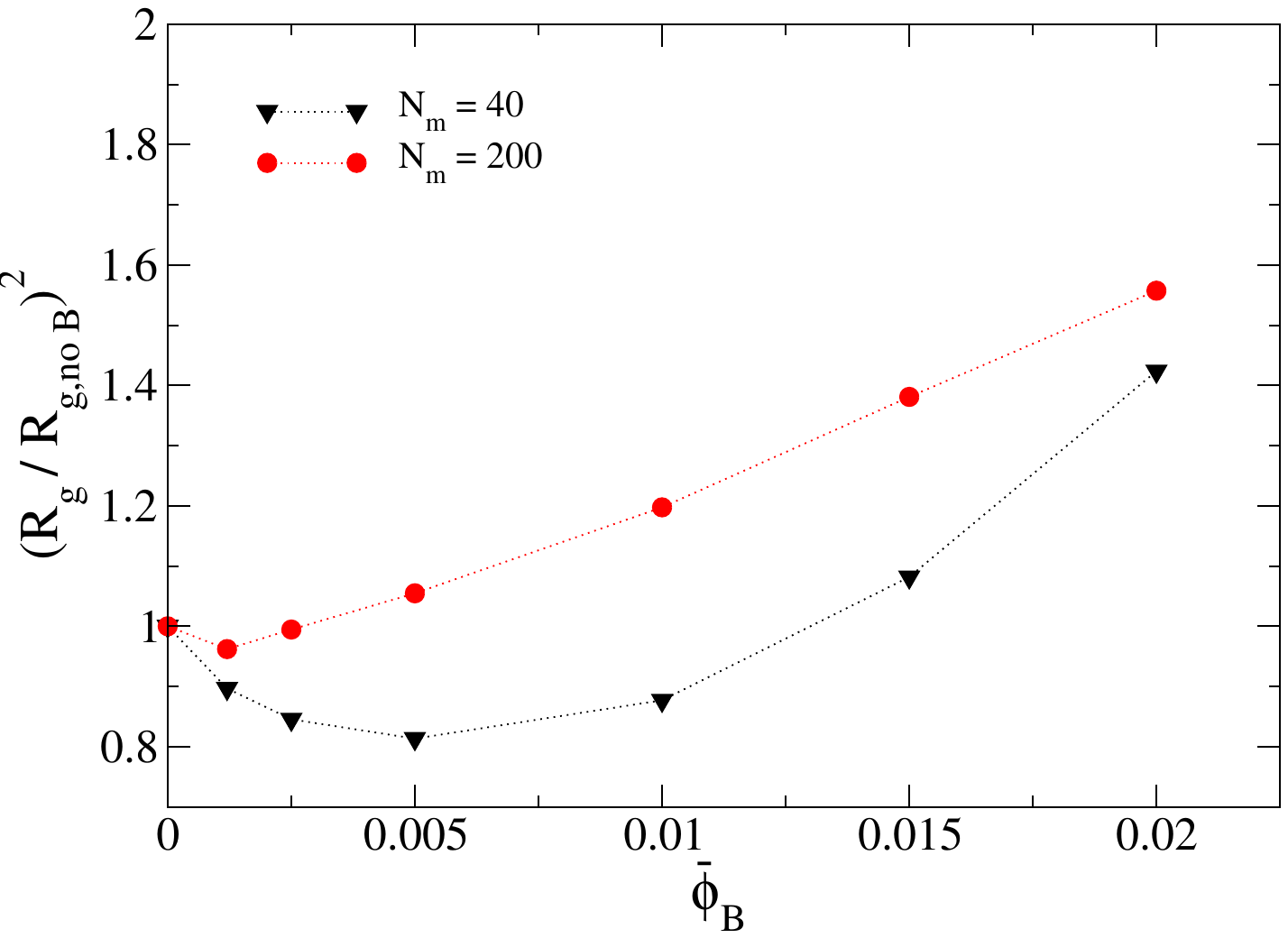}
\caption{Brownian Dynamics : Square of the ratio of the radius of gyration, $R_g$, with a reference value calculated in absence of droplet material, $R_{g, \mathrm{no \ B}} =: R_{g, \bar{\phi}_{B}=0}$, as a function of the average volume fraction of B $\bar{\phi}_{B}$, for different values of the number of monomers $N_{m}$. The Lennard-Jones interaction potentials have energy parameters $\varepsilon_{BB} = 1.2$,  $\varepsilon_{mm} = 0.6 $ and $\varepsilon_{mB} = 1.2$.} 
\label{fig:fig_8b}
\end{center}
\end{figure}

Lastly, we checked whether 3D Brownian Dynamics (BD) simulations of similar polymer systems lead to similar finite size effects. 
In Fig. \ref{fig:fig_8b}, results from BD simulations are shown for chains of 40 and 200 monomers in the presence of B solutes, in a case where (1) B solutes form liquid droplets (2) finite size effects are important in absence of B solutes. We recover the main qualitative results of our minimal model: (1) apart from a shallow minimum at very low B density, the largest polymer monotically expand as we add B particles in the system; (2) the smallest polymer first condenses at low B density, and then, at sufficiently large B density, the chain expands when B density increases.  
The success of these comparisons shows that our minimal model is effective in predicting original qualitative behaviors of polymers coupled to a fluid undergoing LLPS.

\section{\label{sec:concl} conclusions}

In this paper, we have introduced a new minimal model of polymer that can be efficiently coupled to a fluid undergoing liquid-liquid phase transition (LLPS). 
The design of our model was motivated by many recent observations that LLPS might play a role in the nucleoplasm, the fluid filling the nucleus of eukaryotic cells. The formation of droplets of liquids occur in the vicinity of chromatin, the hetereopolymer carrying genetic information~\cite{Gibson_2019,Rippe_2022,Fernandez_2023,mann2023transcription}. Chromatin is compartmentalized in domains with specific epigenetic marks (or epigenetic domains). These domains might be found in a locally swollen or compact configurational state~\cite{Lesage_2019}, a folding variability which may be related to the presence of well defined liquid droplets. 

From the theoretical point of view, a traditional starting point to investigate phase transitions is to start from a mean-field approach~\cite{Berry_2018}. However, an exact mean-field description that couples a multicomponent interacting mixture to a single polymer chain playing the roles, respectively, of the nucleoplasm and an epigenetic domain of chromatin is not straightforward~\cite{Rouches_2024}. In this work, we have investigated such a coupling by means of a minimal approach, describing the system with many interacting particles when one species (the monomer species) is trapped in a harmonic potential. 
The advantage of this approach is to provide a unified theoretical framework that naturally couples ingredients driving the extension of a polymer chain extension to a fluid under LLPS. 

Our first important result concerns the ability of our model to get important qualitative features of interacting polymer chains. We find that a fluid of interacting particles confined by a harmonic potential is a sufficient model to obtain (1) the right extension of the chain in absence of interaction (Gaussian chain); (2) a  dilute and dense phase, similar to a coil (swollen) and a globular (compact) phase of a polymer chain at thermodynamic equilibrium, with a continuous cross-over of both phases when the size of the polymer changes. Previous analysis of epigenetic domain images using on-lattice Monte-Carlo simulations of polymers have already shown the significance of this phenomenon to model chromatin~\cite{Lesage_2019}.  

This minimal model allows to study the influence of a third species, the droplet material (B), on the polymer extension, as quantified by the radius of gyration. The monotonic evolution of the radius of gyration with the number of particles of B can be related to the continuous evolution of solvent quality when B and S mix well together. When S and B fluids demix and form droplets enriched in B particles, and when B particles and monomer attract each other, we have found that the radius of gyration behaves nonmonotically with the density of B particles. 
Such nonmonotonicity shall be related to a phenomenon from the physical chemistry of polymers in solvent mixtures called a co-non-solvency effect~\cite{schild1991cononsolvency,Mukherji_2014}. 
At low volume fraction in B, B particles act as binders, and increasing their concentration result in a condensation of the polymer. However, at larger concentrations, B particles start to act as a good solvent, and the polymer expands.  

Last but not least, the size of the polymer may have a strong influence. For the exact same set of energy parameters, the size of a small polymer may evolve nonmonotically with the quantity of B particles, while a large polymer may expand monotically. All these results have been checked using Brownian dynamics simulations, that provide exact results for interacting polymer models coupled to a fluid of free interacting particles forming droplets. 

Our results suggest new relationships between the material chemistry of chromatin and the surrunding nucleoplasm, and the processes that regulate 3D chromatin domain structure.  
Our study complements other recent works showing coupling between droplets rich in specific proteins and chromatin. From a very different perspective, several groups~\cite{tortora2023hp1,Rouches_2024} have quantified how the polymer favors the phase transition through the existence of \emph{pre-wet phases}. We bring here new 
physics, highlighting how the folding response of a chromatin domain to the recruitment of proteins and the formation of condensates does not only depend on how all the components at play interact with each other, but also on the quantity of droplet material and on the size of the domain. 

Several extensions of this work could be envisioned. First, we could extend our minimal model to include electrostatic interactions. Gradient of electrostatic potential, or indirectly of pH, have been experimentally measured for mesoscale biocondensates including nucleic acids~\cite{ausserwoeger2024biomolecular,welsh2022surface}. In this case, the diffusive currents in the continuity equations of the monomers and of the droplet material need to be modified, e.g. using an extended version of the Flory-Huggins free energy~\cite{Brangwynne_2015}. Next, the droplets may be maintained out-of-equilibrium by chemical reactions~\cite{Laghmach_2021, Zwicker2022}. In the context of nuclear condensates, it is well known that the evolution of the droplets is related to chemical reactions, such as phosphorylation~\cite{nosella2021phosphorylation}. Our model could be extended to reactive systems, allowing to check how non-equilibrium mechanisms may lead to qualitatively different folding states of chromatin domains. Finally, more quantitative estimations of the co-non-solvency effect at chromatin domains will require additional modeling and experimental study, involving extended Brownian Dynamics simulations in combination with the analysis of experimental data. 

\section{\label{sec:sm} supplementary material}


\subsection{\label{sec:sec_1}  Thermodynamics of multicomponent interacting systems in a potential \protect\\}

In this section, we start from the free energy of a system of many interacting components to derive the exchange chemical potentials of the binary and ternary models, Eqs. (3) and (18)-(19) of the main text, respectively.  

The free energy of a multicomponent interacting system of volume $\Omega$ at temperature $T$ fixed by a heat bath generally writes, in $d$ spatial dimensions \cite{Cahn_1958, Berry_2018, Weber_2019, Kirschbaum_2021,Kirschbaum_2022},

\begin{equation}
F = \int d^{d}{r} \left( f (\phi_1, ...\phi_N) -\sum_{i,j}^{N} \frac{\kappa_{ij}}{2v_{ij}}  \nabla \phi_i \nabla \phi_j + U  (\underline{r}) \sum_i  \frac{\phi_i}{v_i}   \right),
\label{eq:eq_1a}
\end{equation}
where $f (\phi_1, ...\phi_N) $ is the homogenous part of the system free energy density which is a function of the volume fractions of all $N$ components ($i = 1, ..., N$), $\kappa_{ij}$ quantifies the energetic cost related to the formation of distinct phases involving the interactions between species $i$ and $j$, and $v_{ij} = l_{ij}^{d}$ is given by the intermolecular distance $l_{ij}$ between species $i$ and $j$. Since $\kappa_{ij}$ arises from a Taylor expansion in terms of gradients of volume fractions of the free energy density around the homogenous part, $f (\phi_1, ...\phi_N)$, $\kappa_{ij}$ might be referred to as the expansion coefficient. The last term in Eq. (\ref{eq:eq_1a}) describes the effect of a (scalar) potential on the volume fraction of all species, where $v_{i} = v_{ii}$ is the molecular volume of species $i$.

A simple form of $f (\phi_1, ...\phi_N) $ is derived from the lattice partition function of an incompressible mixture of identical particles, $v_{ij} = v = v_i$, interacting with their nearest neighbors via pairwise interactions. We then find, in the mean-field limit \cite{Weber_2019, Kirschbaum_2021, Kirschbaum_2022},
\begin{equation}
f (\phi_1, ...\phi_N) = \frac{k_B T}{\nu} \sum_{i} \phi_i \mathrm{ln} \phi_i + \sum_i \frac{\epsilon_i}{v} \phi_i + z\sum_{i,j} \frac{\epsilon_{ij}}{2v} \phi_i \phi_j.
 \label{eq:eq_2a}
\end{equation}
which can also be written as, 
\begin{equation}
	\begin{split}
f (\phi_1, ...\phi_N) & = \frac{k_B T}{\nu} \sum_{i} \phi_i \mathrm{ln} \phi_i + \sum_i \left(\epsilon_i +\frac{z\epsilon_{ii}}{2}\right) \frac{\phi_i}{v} \\
& + {k_BT} \sum_{i,j} \frac{\chi_{ij}}{2v} \phi_i \phi_j,
 \label{eq:eq_3a}
 	\end{split}
\end{equation}
by using the incompressibility condition, $\sum_i \phi_i = 1$, and by defining the Flory-Huggins interaction parameter, $\chi_{ij}$, as,
\begin{equation}
\chi_{ij} =  \frac{z}{2k_B T} \left(2\epsilon_{ij}-\epsilon_{ii}-\epsilon_{jj} \right),
 \label{eq:eq_4a}
\end{equation}
where $z$ is the number of neighbors per lattice site (e.g., $z=4$ and $z=6$ for square and cubic lattices, respectively). In Eq. (\ref{eq:eq_1a}), the expansion coefficient of a binary mixture, $\kappa_{ij} =\kappa$ is related to the Flory-Huggins interaction parameter, $\chi_{ij} =\chi$, by the relation, $\kappa = k_{B}T v^{2/d} \chi$. A simple extension to multicomponent mixtures gives \cite{Kirschbaum_2021, Kirschbaum_2022}, $\kappa_{ij} = k_{B}T v^{2/d} \chi_{ij}$.

From Eq. (\ref{eq:eq_1a}), the chemical potential of a given species $i$, of number $N_i = \phi_i \Omega/v$, is generally expressed as the functional derivative of $F = \int g (\underline{r}, \Phi, \nabla \Phi) d^d{r}$, where $\Phi = (\phi_1, ..., \phi_N)$, which gives,
\begin{equation}
\mu_i = \left(\frac{\delta F}{\delta N_i}\right)_{N_{j\ne i},\Omega,T} = v \left( \frac{\partial g}{\partial \phi_i} - \nabla \frac{\partial g}{\partial \nabla \phi_i} \right)_{N_{j\ne i},\Omega,T},
 \label{eq:eq_5a}
\end{equation}
where $g (\underline{r}, \Phi, \nabla \Phi)$ is the total free energy density which contains all the terms in the bracket of Eq. (\ref{eq:eq_1a}). 

We now apply the relations for a ternary system composed of the monomer (m), droplet material (B), and solvent (S) species. Let us assume only the monomers to be affected by the presence of the potential, $U$. The total free energy density then writes, 

\begin{equation}
	\begin{split}
& \frac{v}{k_BT}	g (\underline{r}, \Phi, \nabla \Phi) =  \phi_m \mathrm{ln} \phi_m +  \phi_B \mathrm{ln} \phi_B + \phi_S  \mathrm{ln} (\phi_S) \\
	& +\frac{1}{k_BT} \left(w_m \phi_m + w_B \phi_B + w_S \phi_S\right) + \chi_{mB} \phi_{m} \phi_{B} +  \chi_{mS} \phi_{m} \phi_{S}   \\
	& + \chi_{BS} \phi_{B} \phi_{S} - v^{2/d} \biggl(\chi_{mB} \nabla {\phi_m} \nabla {\phi_B} + \chi_{mS} \nabla {\phi_m} \nabla {\phi_S}  \\
	& +  \chi_{BS} \nabla {\phi_B} \nabla {\phi_S} \biggr) + \frac{U}{k_{\mathrm{B}}T} \phi_m,
\label{eq:eq_6a}
\end{split}
\end{equation}  
where $\Phi=(\phi_m, \phi_B, \phi_S)$, $\nabla \Phi = (\nabla \phi_m, \nabla \phi_B, \nabla \phi_S)$, and $w_i = \epsilon_i +z\epsilon_{ii}/2$ (with $i = m, B$ or $S$).

From Eqs. (\ref{eq:eq_5a})-(\ref{eq:eq_6a}), the two exchange chemical potentials, $\mu = \mu_m - \mu_S$ and $\mu' = \mu_B - \mu_S$, whose gradients provide the driving forces of diffusion processes of the monomer and of the droplet material species, respectively, then become,

\begin{equation}
	\begin{split}
	&\mu = {w_m - w_S} +k_{\mathrm{B}}T \mathrm{ln}\frac{\phi_m}{1-\phi_m-\phi_B}   \\
	& \ + k_{\mathrm{B}}T (\chi_{mB} \textcolor{black}{-\chi_{BS}}) \phi_B + k_{\mathrm{B}}T \chi_{mS} (1-2 \phi_m - \phi_B) \\
	& \  + \bar{\kappa} \nabla^2 \phi_B - {\kappa_{mS}} \nabla^2 \phi_m +U,
\label{eq:eq_7a}
\end{split}
\end{equation}

\begin{equation}
	\begin{split}
	&\mu^{\prime} = {w_B - w_S} +k_{\mathrm{B}}T \mathrm{ln}\frac{\phi_B}{1-\phi_m-\phi_B}  \\
	& \ + k_{\mathrm{B}}T (\chi_{mB} \textcolor{black}{-\chi_{mS}}) \phi_m + k_{\mathrm{B}}T \chi_{BS} (1-2 \phi_B - \phi_m)   \\
	& \  + \bar{\kappa} \nabla^2 \phi_m - {\kappa_{BS}} \nabla^2 \phi_B ,
\label{eq:eq_8a}
\end{split}
\end{equation}
where $\bar{\kappa} = k_{\mathrm{B}}T  v^{2/d} \left( \chi_{mB}-\chi_{BS}-\chi_{mS}\right)$, and $\kappa_{ij} =  k_{\mathrm{B}}T  v^{2/d} \chi_{ij}$. 

Eqs. (\ref{eq:eq_7a})-(\ref{eq:eq_8a}) respectively recover Eqs. (18)-(19) of the main text. In the absence of B ($\phi_B = 0$), Eq. (\ref{eq:eq_7a}) becomes,

\begin{equation}
	\begin{split}
\mu &= {w_m - w_S} + k_{\mathrm{B}}T \mathrm{ln}\frac{\phi_m}{1-\phi_m} + k_{\mathrm{B}}T  \chi (1-2\phi_m) \\
&\ -\kappa \nabla^2 \phi_m +U,
\label{eq:eq_9a}
\end{split}
\end{equation}
where $\kappa = \kappa_{mS}$. Eq. (\ref{eq:eq_9a}) recovers Eq. (3) of the main text.


\subsection{\label{sec:binary_1} Equilibrium solutions when $U=0$}

We recall here the classical model of liquid-liquid phase separation of a binary mixture in the absence of any external force field or potential. 

In the long-time limit, the volume fraction at each location relaxes to its equilibrium value, denoted by $\phi = \phi_m (\underline{r}, t\rightarrow \infty)$. Equilibrium volume fractions minimize the system free energy $F$ under the constraints of mass and volume conservation. From Eq. \ref{eq:eq_1a}, the system free energy of a binary mixture in the absence of the potential ($U=0$) writes \cite{Weber_2019}:

\begin{equation}
F = \int d^{d}{r} \left( f (\phi) + \frac{\kappa}{2v} \lvert \nabla \phi \lvert^2  \right),
\label{eq:eq_7}
\end{equation}
where the free energy density reads
\begin{equation}
	\begin{split}
f (\phi) = \frac{w_m}{v} \phi + \frac{w_S}{v} (1- \phi) + \frac{k_{\mathrm{B}}T}{v} f_{mix},
\label{eq:eq_8}
\end{split}
\end{equation}
where $ f_{mix} (\phi)$ is the free energy density of \textit{mixing} \cite{qian2022analytical}:
\begin{equation}
f_{mix} (\phi)= \phi \mathrm{ln} \phi + (1-\phi) \mathrm{ln} (1-\phi) + \chi \phi(1-\phi),
\label{eq:eq_8b}
\end{equation}
which separates the total free energy density from the internal contributions. 

A homogeneous mixture is locally unstable when $f''(\phi) < 0$, and thus the \textit{spinodal} boundary is defined by $f''(\phi) = 0$ \cite{Weber_2019, qian2022analytical}. From the latter condition, we find two solutions for the \textit{spinodal} volume fractions related to a dense ($\phi^{\mathrm{spin}}_{+}$) and a dilute phase ($\phi^{\mathrm{spin}}_{-}$), that write, 
\begin{equation}
\phi^{\mathrm{spin}}_{\pm} = \left( \frac{1}{2} \pm \sqrt{\frac{1}{4}-\frac{1}{2\chi}} \right).
\label{eq:eq_9}
\end{equation}

The \textit{binodal} volume fractions are obtained by  minimization of the total free energy, $F = \Omega_{+} f(\phi_+) + \Omega_{-} f(\phi_-)$, performed under the constraints of mass ($\Omega_{+} \phi_+ + \Omega_{-} \phi_- = \Omega \bar{\phi}$, where $\bar{\phi}$ denotes the spatial average of $\phi$) and volume conservation ($\Omega = \Omega_+ + \Omega_-$), where $\phi_+$ and $\phi_-$ denote the \textit{binodal} volume fractions of the two phases of volumes $\Omega_+$ and $\Omega_{-}$ , respectively \cite{Weber_2019, qian2022analytical}. This gives the two following conditions:

\begin{equation}
 f'(\phi_{-}) =  f'(\phi_{+}),
\label{eq:eq_10a}
\end{equation}

\begin{equation}
f(\phi_{-}) - f(\phi_{+}) + \left(\phi_{+}-\phi_{-} \right) f'(\phi_{+}) = 0.
\label{eq:eq_10b}
\end{equation}
Eqs. \ref{eq:eq_10a}-\ref{eq:eq_10b} denote the balance of the exchange chemical potentials and of the osmotic pressures between the two phases, respectively. 

Equilibrium solutions, $\phi_{-}$ and $\phi_{+}$, satisfying Eqs. \ref{eq:eq_10a}-\ref{eq:eq_10b} can be found graphically by a common tangent construction. The general solution reads \cite{Weber_2019, qian2022analytical},

\begin{equation}
\chi = \biggl[ \frac{\mathrm{ln} \left({\phi}/{1-\phi}\right)}{(2\phi-1)} \biggr]_{\mathrm{bin}},
\label{eq:eq_11}
\end{equation}
which is unique when $(\phi, \chi) = (\phi_c, \chi_c)$, which defines the critical point of LLPS, $\phi_c =1/2$ and $\chi_c = 2$. Consequently, phase separation is only possible when $\chi > \chi_c=2$.

\subsection{\label{sec:sec_2}  Equilibrium solutions of the binary and ternary models \protect\\}

The goal of this section is to recover the equilibrium solution of the binary model, given by Eq. (10) of the main text. We also extend this result for completeness to the ternary case.

The binary model, in dimensionless form, reads (Eq. (7) of the main text),

\begin{equation}
	\frac{\partial \phi_m}{\partial t} (\underline{r},t) = \nabla \, \left[ \phi_m \nabla \mu _{U=0} + \phi_m \nabla U \right].
\label{eq:eq_1}
\end{equation}

At equilibrium, $({\partial \phi_m}/{\partial t}) = 0$ and the two terms in Eq. (\ref{eq:eq_1}) balance each other, which gives, the \textit{equilibrium} condition,
\begin{equation}
	\ \nabla \left[ \mu _{U=0} + U \right] = 0.
\label{eq:eq_2}
\end{equation}
Since the equilibrium condition, Eq.~(\ref{eq:eq_2}), is a condition on the \textit{gradient} of the chemical potential, any term independent on space in the chemical potential do not contribute to the monomer volume fraction solutions and thus, can be dropped hereafter.

From Eqs. (3)-(4) (with $k_{\mathrm{B}}T = 1$), it follows that the equilibrium monomer volume fraction, $\phi_m (\underline{r},t\rightarrow \infty) = \phi$, must satisfy the relation,
\begin{equation}
	\begin{split}
 \biggl[ \mathrm{ln}\frac{\phi}{1-\phi} -2\phi  \chi -\chi \nabla^2 \phi +U \biggr]   = C_1, \\
&\ ,
\label{eq:eq_3}
\end{split}
\end{equation}
where $C_1$ is a non-zero constant (in space and time). From Eq. (\ref{eq:eq_3}), we find, 

\begin{equation}
	\begin{split}
  \frac{\phi}{1-\phi}  = C_2 \mathrm{exp} \biggl(-U \biggr)  \mathrm{exp} \biggl(2 \chi \phi +\chi \nabla^2 \phi  \biggr), \\
&\ ,
\label{eq:eq_4}
\end{split}
\end{equation}
where $C_2 = \mathrm{exp}{(C_1)}$. In Eq. (\ref{eq:eq_4}), $C_2$ can be written in an integral form as,

\begin{equation}
	\begin{split}
C_2 = \frac{\langle \phi/(1-\phi)\rangle_{x,z}}{\langle  \mathrm{exp} (-U) \times \mathrm{exp}  \left(2\phi\chi + \chi \nabla^2 \phi \right) \rangle}_{x,z}, \\
&\ ,
\label{eq:eq_5}
\end{split}
\end{equation}
where $\langle f\rangle_{x,z}$ denotes the surface average of a scalar function $f$, given by $\langle f \rangle_{x,z} = \int_{0}^{L} \int_{0}^{L} f(x,z) dxdz/L^2$. Eq. (\ref{eq:eq_5}) is obtained after integration over the entire (square) domain $\Omega = \ [0,L] \times [0,L]$ of the left and right sides of Eq. (\ref{eq:eq_4}).

From Eq. (\ref{eq:eq_5}) and since $U= \alpha \lvert  \underline{r}-\underline{r}_c  \lvert^2 $, Eq. (\ref{eq:eq_4}) becomes,

\begin{equation}
	\begin{split}
	&\frac{\phi/(1-\phi)}{\langle \phi/(1-\phi) \rangle_{x,z}} (\underline{r}) = \\
	&\ \frac{ \mathrm{exp}  \left(-\alpha \lvert  \underline{r}-\underline{r}_c \lvert^2\right)  \times \mathrm{exp}  \left(2\phi\chi + \chi \nabla^2 \phi \right)}{\langle \mathrm{exp}  \left(-\alpha \lvert  \underline{r}-\underline{r}_c \lvert^2\right)  \times \mathrm{exp}  \left(2\phi\chi + \chi \nabla^2 \phi \right) \rangle_{x,z}},  \\
\label{eq:eq_6}
\end{split}
\end{equation}
which recovers Eq. (10) of the main text.

Similarly, for the ternary model, defined by Eqs. (14)-(15) of the main text, the equilibrium solution for the monomer volume fraction must satisfy the relation,

\begin{equation}
	\begin{split}
	&\frac{\phi_m/(1-\phi_m-\phi_B)}{\langle \phi_m/(1-\phi_m-\phi_B) \rangle_{x,z}} (\underline{r}) = \\
	&\ \frac{ \mathrm{exp}  \left(-\alpha \lvert  \underline{r}-\underline{r}_c \lvert^2\right)  \times \mathrm{exp}  \left(\mathcal{-B} \right)}{\langle \mathrm{exp}  \left(-\alpha \lvert  \underline{r}-\underline{r}_c \lvert^2\right)  \times \mathrm{exp}  \left(\mathcal{-B}  \right) \rangle_{x,z}},  \\
 \label{eq:eq_7}
\end{split}
\end{equation}
where $\phi_m$ and  $\phi_B$ here denote the equilibrium volume fractions of the monomers and of the droplet material, and $\mathcal{B} =: \mathcal{B} (\phi_m, \phi_B, \nabla^2 \phi_m, \nabla^2 \phi_B)$ is defined as,

\begin{equation}
	\begin{split}
\mathcal{B} &=  (\chi_{mB} \textcolor{black}{-\chi_{BS}}) \phi_B - \chi_{mS} (2\phi_m + \phi_B) \\
&\ + \left( \chi_{mB}-\chi_{BS}-\chi_{mS}\right)\nabla^2 \phi_B - {\chi_{mS}} \nabla^2 \phi_m.
 \label{eq:eq_8}
 \end{split}
\end{equation}

Eqs.~\ref{eq:eq_6}-\ref{eq:eq_7} are of interest to check the accuracy of the numerical integration as performed in the finite-element based COMSOL Multiphysics software \cite{COMSOL_2021}.

\subsection{\label{sec:sec_3}  Total number of particles of a given species and its conservation in time \protect\\}

Here, we show that the total number of a given species $i$ is given by, $N_i = \bar{\phi}_i L^2$ (in dimensionless form), where $\bar{\phi}_i $ is the surface average of $\phi_i$ (where $i$ = m, S or B), and is a quantity which is conserved in time.

Both the binary and ternary models are built from continuity equations of the form (Eq. (1) and Eqs. (14)-(15) of the main text),
\begin{equation}
	\frac{\partial \phi_{i}}{\partial t} (\underline{r},t)  = - \nabla \, \underline{j}^{\mathrm{dif}}_{i} (\underline{r},t),
\label{eq:eq_9}
\end{equation}
for a species $i$.
By integrating the left and right sides of Eq. (\ref{eq:eq_9}) over the entire domain, $\Omega$, in $d$ spatial dimensions,

\begin{equation}
	\frac{\partial}{\partial t} \int_{\Omega} \phi_i  d^d{r}  = - \int_{\Omega} \nabla \, \underline{j}^{\mathrm{dif}}_{i}   d^d{r},
\label{eq:eq_10}
\end{equation}
where $\int_{\Omega} \phi_m d\underline{r} = N_m$ is the total number of monomers. According to the Gauss' theorem, the right-hand side of Eq. (\ref{eq:eq_9}) is the integral over the boundary $\partial \Omega$ of the vector field (diffusion flux, $\underline{j}^{\mathrm{dif}}_{i}$), 

\begin{equation}
\int_{\Omega} \nabla \, \underline{j}^{\mathrm{dif}}_{i}  d^d{r} = \int_{\partial \Omega} \, (\underline{j}^{\mathrm{dif}}_{i} \underline{n})  d r^{d-1},
\label{eq:eq_11}
\end{equation}
where $\underline{n}$ is the vector normal to the boundary, $\partial \Omega$. In a closed system (no matter going inside or outside the system through the system boundaries), Eq. \ref{eq:eq_11} vanishes. Hence, from Eq. \ref{eq:eq_10}, we find that (in dimensionless form), 

\begin{equation}
 \int_{\Omega} \phi_i d^dr =: N_i,
 \label{eq:eq_12}
\end{equation}
is conserved in time. Then, Eq. (\ref{eq:eq_12}) can be evaluated at the initial time, $t=0$, to give, 

\begin{equation}
\bar{\phi}_i L^2 =: N_i,
 \label{eq:eq_13}
\end{equation}
in $d=2$ dimensions on the domain of interest, $\Omega = \ [0,L] \times [0,L]$, and where, $\bar{\phi}_i =: {\phi}_i (t=0)$. From Eq. \ref{eq:eq_13}, we note that $\bar{\phi}_i $ is also the surface average of $\phi_i$, 

\begin{equation}
\bar{\phi}_i = \frac{1}{L^2} \int_{0}^{L} \int_{0}^{L} \phi_i dxdz.
\label{eq:eq_14}
\end{equation}




\subsection{\label{sec:sec_4}  Brownian dynamics: simulations details \protect\\}
Brownian dynamics simulations based on an overdamped Langevin
equation are used to simulate a polymer chain embedded in a solvent in the presence of free solute particles B.
More precisely, the
displacement of $N$ particles included in the simulation box from
time $t$ to time $t+\delta t$ reads {\cite{ErmakJCP75}}:
\begin{equation}
\label{ermak} {\bf r}(t+\delta t) = {\bf r}(t) +  
{ D^\circ}{} {\bf F}(t) \delta t + \sqrt{2D^\circ\Delta t}{\bf R},
\end{equation}
\noindent where $D^\circ$ is the self--diffusion coefficient of 
particles at infinite dilution, $\delta t$ is the time increment,
${\bf r}$ is the $3N$-dimensional configuration vector, and ${\bf F}$ is the total force acting on the particles at the beginning of
the step, deriving from a potential. $\bf R$ is a random displacement, chosen from a Gaussian
distribution with zero mean, $\langle {\bf R}\rangle ={\bf 0}$, and
variance $\langle {\bf R} {\bf R}^T \rangle = 
{\bf I}$, the identity matrix.  Here, we use dimensionless quantities, with the diameter of monomers and B particles $\sigma$ as the unit of length, the thermal energy  $k_{\rm B}T$ with $k_{\rm B}$ the Boltzmann constant and $T$ the temperature as the unit of energy, and the typical time taken by a particle to diffuse over its own diameter $\frac{\sigma^2}{D^\circ}$ as the unit of time. The classical overdamped Langevin algorithm used in BD simulations is quite unstable for polymers, unless the time steps are considerably smaller than the one used for LJ fluids. More efficient algorithms have been developed, without affecting the ability to compute structural properties.  We used here a Metropolised Langevin Algorithm~\cite{HeyesMolPhys98, MALA,Rossky78,
JardatJCP99}, implemented in an in-house code written in Fortran.

The interaction pair potential between bonded monomers is the sum of two terms. First, we have the Finitely Extensible Nonlinear Elastic (FENE) potential:
\begin{equation}
V_{\rm FENE}( r )=-\frac{1}{2}k_{\rm FENE}R_0^2\ln \left( 1-\frac{r^2}{R_0^2}\right)
\end{equation}
where $k_{\rm FENE}$ is the spring constant and $R_0$ is the maximum
extension of the bond. 
Second, we add a Weeks-Chandler-Andersen short-range interaction potential
\begin{equation}
\label{eq:def_WCA}
    U_{\text{WCA}}(r) =
    \begin{cases}
    4\varepsilon'\left[ \left( \frac{\sigma}{r}\right)^{12}-\left(\frac{\sigma}{r} \right)^{6}\right] + \varepsilon' & \text{if $r<2^{1/6}\sigma$}, \\
    0 & \text{otherwise}.
    \end{cases}
\end{equation}
The parameters of these interaction potentials for bonded monomers are: $\sigma=1$, $k_{\rm FENE}=7$, $R_0=2$, $\varepsilon'=0.833$. With these values, the distance between bonded monomers is equal to $\sigma$.

Interactions between non-bonded monomers, between B particles and between monomers and B particles are described by a Lennard-Jones (LJ) interaction potential:
\begin{equation}
     U_{\text{LJ}}(r) =
     \begin{cases}
    4\varepsilon\left[ \left( \frac{\sigma}{r}\right)^{12}-\left(\frac{\sigma}{r} \right)^{6}\right] & \text{if $r<2.5\sigma$}, \\
    0 & \text{otherwise}.
    \end{cases}
\end{equation}
We take $\sigma=1$ both for monomers and B particles. The values of $\varepsilon$ depends on the particles interacting with each other: they are denoted by $\varepsilon_{mm}$ for the LJ between non-bonded monomers, $\varepsilon_{BB}$ for the LJ between B particles, and $\varepsilon_{mB}$ for the LJ between B and monomers. 

For each system investigated, $1$ polymer chain is put in a cubic simulation box together with particles B at a random initial configuration of B particles. The initial configuration of the polymer is a random coil. First, long simulations are performed to ensure that equilibrium is reached. As an indicator of the convergence to equilibrium, we follow the temporal evolution of the gyration radius of the chain $R_g$, as well as the temporal evolution of the radial distribution functions between B particles. The gyration radius is computed as
\begin{equation} \label{eq:RG1}
	R_{\rm g}^2
	    = \frac{1}{N}\sum_{i=1}^{N} \left( {\bf r}_{i} - {\bf  r}_G \right)^2
	    = \frac{1}{N}\sum_{i=1}^{N} ({\bf r}_i^2 - {\bf  r}_G^2),
	\quad \mathrm{with} \quad {\bf  r}_G = \frac{1}{N}\sum_{i=1}^{N} {\bf  r}_i \,
\end{equation}
$N$ being the number of monomers of the chain, and ${\bf r}_i$ the position of the $i$-th monomer. Then, several independent initial configurations representative of the equilibrium are propagated with BD for $N_{prod}$ time steps.
The time step is equal to $10^{-5}\frac{\sigma^2}{D^\circ}$ in every case. It allows to keep an acceptance rate of the MALA algorithm larger than $0.8$. For each systems, results are averaged over $10$ to $50$ independent trajectories, each one with between $5\times 10^6$ to $5 \times 10^8$ steps, depending on the total density of the system. The number of monomers varies between $20$ and $200$, and the amount of B particles varies between $0$ and $4000$. The length of the simulation box is kept fixed to $29.7\sigma$, so that the mean volume fraction of B particles $\bar\phi_B$ varies between $0$ and $0.08$. The LJ parameters between monomers $\varepsilon_{mm}$ are either $0.2$ or $0.6$. The LJ parameters for the interaction between B particles  $\varepsilon_{BB}$ is chosen close to the value where a liquid-gaz transition appears at the chosen density: it is either $1.0$ or $1.2$. In every case, we take  $\varepsilon_{mB}=\varepsilon_{BB}$ for simplicity.

\section*{acknowledgements}
R.T. acknowledges the ``Agence Nationale de la Recherche$"$ (ANR) for their financial support (ANR project ANR-19-CE45-0016.).

\nocite{*}
\bibliography{JCP_v1}

\end{document}